\documentclass[iop, twocolumn, tighten, twocolappendix]{emulateapj}
\pdfoutput=1
\usepackage{amsmath}

\usepackage{natbib}

\usepackage[backref,breaklinks,colorlinks,allcolors=blue]{hyperref}

\usepackage{xspace}
\newcommand{\kms}{km~s$^{-1}$\xspace}

\shorttitle{Compaction of primitive solar system solids}
\shortauthors{Davison et al.}

\slugcomment{Submitted to ApJ on Dec.03, 2015; Revised on Feb.11, 2016; Accepted on Feb.25, 2016}
\begin{document}

\title{Mesoscale Modeling of Impact Compaction of Primitive Solar System Solids}

\author{Thomas M. Davison\altaffilmark{1} Gareth S. Collins\altaffilmark{1} and Philip A. Bland\altaffilmark{2}}
\affil{$^{1}$Impacts and Astromaterials Research Centre,
    Department of Earth Science and Engineering,
    Imperial College London, South Kensington Campus,
    London SW7 2AZ, UK;\\
    \href{mailto:thomas.davison@imperial.ac.uk}{thomas.davison@imperial.ac.uk}\\
    $^{2}$Department of Applied Geology, Curtin University of Technology, 
    GPO Box U1987, 
    Perth WA 6845, Australia}

\begin{abstract}
We have developed a method for simulating the mesoscale compaction of early solar system solids in low velocity impact events, using the iSALE shock physics code. Chondrules are represented by nonporous disks, placed within a porous matrix. By simulating impacts into bimodal mixtures over a wide range of parameter space (including the chondrule-to-matrix ratio, the matrix porosity and composition and the impact velocity), we have shown how each of these parameters influences the shock processing of heterogeneous materials. The temperature after shock processing shows a strong dichotomy: matrix temperatures are elevated much higher than the chondrules, which remain largely cold. Chondrules can protect some matrix from shock compaction, with shadow regions in the lee side of chondrules exhibiting higher porosity that elsewhere in the matrix. Using the results from this mesoscale modelling, we show how the $\varepsilon-\alpha$ porous compaction model parameters depend on initial bulk porosity. We also show that the timescale for the temperature dichotomy to equilibrate is highly dependent on the porosity of the matrix after the shock, and will be on the order of seconds for matrix porosities of less than 0.1, and on the order of 10's to 100's seconds for matrix porosities of $\sim$ 0.3--0.5. Finally, we have shown that the composition of the post-shock material is able to match the bulk porosity and chondrule-to-matrix ratios of meteorite groups such as carbonaceous chondrites and unequilibrated ordinary chondrites.
\end{abstract}

\keywords{meteorites, meteors, meteoroids --- methods: numerical --- minor planets, asteroids: general --- planets and satellites: formation --- protoplanetary disks --- shock waves}

\section{INTRODUCTION}
Primitive solar system solids are expected to have accumulated as bimodal mixtures of mm-scale zero-porosity inclusions (chondrules) surrounded by highly porous, sub-$\mu$m dust particles (matrix). Previous numerical simulations of impact processing (e.g. compaction and heating) of such materials have treated the mixture as homogeneous and estimated impact-generated ‚Äòbulk‚Äô shock pressures and temperatures over large (i.e. planetesimal) scales \citep[e.g.][]{Keil:97, Davison:10a}. To model the bimodal mixtures explicitly, and resolve  shock response at the scale of individual chondrules, requires a different numerical approach known as mesoscale modeling \citep[e.g., ][]{Nesterenko:01}. By adopting this approach, \citet{Bland:14a} revealed new insight into the heterogeneous response of chondritic precuror material to impact-induced compaction. 

\citet{Williamson:86} first introduced ``microlevel numerical modeling'' and showed how a shock wave affects a small unit cell of closest packed cylinders of stainless steel, with void or air in the interstitial spaces; that work confirmed the experimental result that there is a concentration of heating at the particle boundaries due to the localisation of plastic deformation. \citet{Williamson:89} extended this model to include a second material. However, this work still only simulated a small cell, which could not provide any information about the material on a larger scale. To extend this type of modeling to investigate the bulk effects of shock waves on heterogenous materials, Eulerian finite element simulations of impacts into randomly packed particles were developed \citep{Benson:94, Benson:97}, and were found to reproduce averaged values of pressure, density and porosity in close agreement with stationary shock experiments into granular materials. This approach has now become widely adopted for characterising the shock response of granular materials \citep[e.g.][]{Borg:08,Borg:12}, including constructing bulk-material Hugoniot relationships and equations of state.

Mesoscale modeling has also been used for planetary impact applications. Simulations of impact crater growth on asteroids have investigated the effects of target grain size and heterogeneous materials \citep{Barnouin-Jha:02, Crawford:03}. The consequences of the presence of water ice on Mars has been investigated by modeling rock/ice mixtures using a range of geometries, including modeling ice inclusions within a rock matrix \citep{Ivanov:05, Ivanov:11}. More recently, \citet{Gueldemeister:13} simulated the mesoscale response of porous and water-saturated materials and found good agreement with both macroscale models (where the porosity was parameterised and the water-saturated material was described by a mixed-material equation of state) and Hugoniot data from shock experiments, for both regularly and randomly distributed pores. Finally,  \citet{Bland:14a} applied mesoscale modeling techniques to the scenario of shock processing of primitive materials, by explicitly modeling shock wave propagation through a mixture of non-porous chondrules distributed within a porous matrix. That work showed that the heterogeneous nature of primitive meteoritic material leads to strong dichotomies in the temperatures experienced by the two different components: Porous matrix was heated to $>$~1000~K by impact velocities of $\sim$~1.5~\kms, while the non-porous chondrules were only heated by tens of Kelvin. Here, we describe the modeling techniques used by \citet{Bland:14a} in more detail, including a sensitivity analysis, and expand the parameter space of the simulations presented in that work.

Using the results of this mesoscale modeling, we then determine: (1) how the parameters of the $\varepsilon-\alpha$ porous compaction model \citep{Wuennemann:06,Collins:11a} depend on the initial bulk porosity for different chondrule-matrix mixtures; (2) the timescale for equilibration of the temperature between the matrix and the chondrules; and (3) that the porosities and chondrule abundances of the post-shock material are able to match those of chondritic meteorite groups.

\section{METHODS}\label{Sect:Methods}
\subsection{The iSALE Shock Physics Code}
To quantify the compaction of porous meteoritic material in an impact we used the iSALE shock physics code \citep{Collins:04, Wuennemann:06}, a multi-material, multi-rheology extension of the SALE hydrocode \citep{Amsden:80}. iSALE incorporates several additions to the original SALE code, including an elasto-plastic constitutive model, fragmentation models, various equations of state (EoS), and multiple materials \citep{Melosh:92,Ivanov:97}. Recent additions include a modified strength model \citep{Collins:04} and a porosity compaction model \citep{Wuennemann:06, Collins:11a}. iSALE has been benchmarked against other hydrocodes \citep{Pierazzo:08} and validated against laboratory impact experiments \citep[e.g.][]{Pierazzo:08, Davison:11a} for crater formation applications. In \cite{Bland:14}, we performed a suite of two-dimensional (2D) plane-strain, mesoscale simulations of shock wave propagation through a bimodal mixture of non-porous chondrules (represented as 2D disks) surrounded by a highly porous continuous matrix. Here we describe and justify the methodology used in that work and supplement those results with several additional simulation suites to expand the physical parameter space of our results and to demonstrate the sensitivity of results to model choices. 

Chondritic meteorites contain (nominally) non-porous spherical chondrules ($\sim$~0.1--1~mm in diameter) surrounded by a highly porous matrix composed of sub-$\mu$m particles. Because the length scale of these two components is so different (approximately three orders of magnitude), we chose to simulate the bimodal mixture by explicitly resolving the chondrules as non-porous (2D) disks, and modeling the porous matrix as a continuum. The compaction of the matrix porosity was computed using the $\varepsilon-\alpha$ porous-compaction model \citep{Wuennemann:06, Collins:11a}, because the length-scale of the porosity implied that it was too small to be resolved in the simulation. ANEOS-derived tabular equations of state were used to describe the chondrules and the matrix. In all cases, the ANEOS table for dunite/forsterite \citep{Benz:89} was used for the chondrules. The solid component of the matrix was described by either the dunite/forsterite ANEOS table, or an ANEOS table for serpentine, created using parameters listed in \cite{Brookshaw:98}. The response of the chondrules and matrix to changes in deviatoric stress was calculated using a geologic strength model \citep{Collins:04}. Chondrules were assigned a high cohesive strength of 1~GPa, while the matrix was assumed to be very weak (cohesive strength of 0.1~MPa; see Table~\ref{Tab:Material} for all the iSALE material input parameters used in this work).

\begin{table}[!tb]
    \begin{center}
    \caption{Material parameters used in numerical simulations}
    \label{Tab:Material}
    \begin{tabular}{lll}
    \hline\hline
    Parameter & Chondrules & Matrix \\
    \hline
    Initial porosity                                           & 0    & 0.7\\
    Compaction rate\tablenotemark{a}                           & N/A  & 0.98\\
    Vol. strain at onset\\
    \ \ of plastic compaction\tablenotemark{a}                 & N/A  & $-10^{-5}$\\
    Poisson ratio (solid component)\tablenotemark{b}           & 0.23 & 0.23\\
    Intact cohesion\tablenotemark{b} (MPa)                     & 1000 & 0.1\\
    Intact friction coefficient\tablenotemark{b}               & 1.2  & 1.2\\
    Intact strength limit\tablenotemark{b} (GPa)               & 3.5  & 0.035\\
    Damaged cohesion\tablenotemark{b} (MPa)                    & 0.01 & 0.01\\
    Damaged friction coefficient\tablenotemark{b}              & 0.6  & 0.6\\
    Damaged strength limit\tablenotemark{b} (GPa)              & 3.5  & 0.035\\
    Melt temperature                                           & 1373\tablenotemark{c,d} & Dunite: 1373\\
    \ \ (zero pressure) (K)                                    &      & Serpentine: 1098\tablenotemark{e,f}\\
    Simon approx. constant\tablenotemark{g} (GPa)              & 1.52 & 1.52\\
    Simon approx. exponent\tablenotemark{g}                    & 4.05 & 4.05\\
    Thermal softening parameter\tablenotemark{b}               & 1.2  & 1.2\\
    \hline
    \end{tabular}
    \tablenotetext{1}{\citet{Wuennemann:06}}
    \tablenotetext{2}{\citet{Collins:04}}
    \tablenotetext{3}{\citet{Keil:97}}
    \tablenotetext{4}{\citet{Katz:03}}
    \tablenotetext{5}{\citet{ATSDR:01}}
    \tablenotetext{6}{\citet{Till:11}}
    \tablenotetext{7}{\citet{Wuennemann:08}}
    \end{center}
\end{table}

\subsection{Simulation Design}
The bimodal mixture of chondrules and matrix was generated by randomly placing circular disks of the chondrule material throughout the computational mesh. The diameters of the disks were randomly distributed in the size range 0.3--1~mm, based on typical chondrule sizes \citep{Scott:03}. Chondrules were added in this way until the desired chondrule-to-matrix volume ratio was reached, and the interstitial space was filled with matrix material. Chondrule-to-matrix fraction was quantified by matrix abundance, $A_{mi}$, the initial matrix volume as a percentage of the total; $A_{mi}$ varied between 30\% and 70\%, which was found \emph{a posteriori} to account for the observed range in current matrix abundance observed in chondritic meteorites \citep{Scott:03}. The initial matrix porosity was typically 0.7, although some simulations examined the effect of changing this to 0.6 or 0.8. The initial temperature was 300~K in all simulations presented here; some test simulations were run with different starting temperatures (down to 170~K), but as these simulations resulted in the same increase in temperature as those reported here, for brevity, only 300~K simulations are discussed below and only absolute temperatures are reported.

\begin{figure*}[t]
    \begin{center}
        \includegraphics[width=\textwidth]{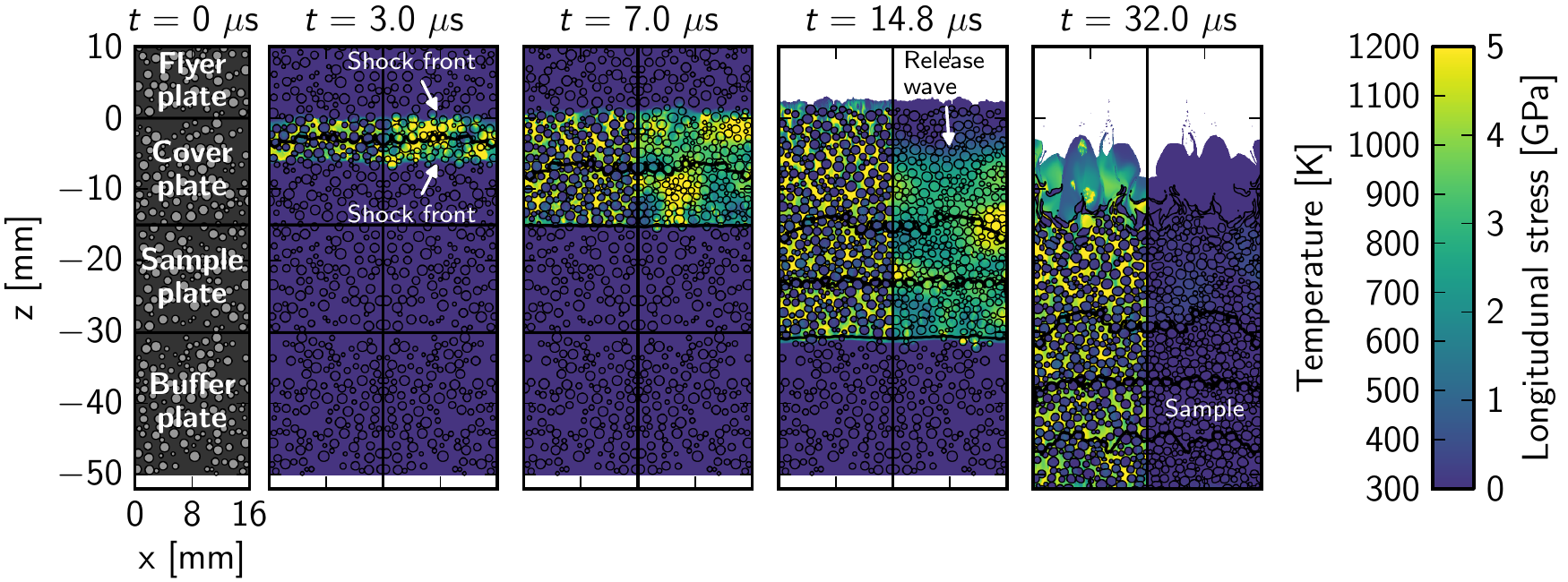}
    \end{center}
    \caption{Time sequence (from left to right) of the reference mesoscale simulation of a nominally planar shockwave propagating through a bimodal mixture of explicitly resolved non-porous chondrules surrounded by a high-porosity matrix ($A_{mi} = 70\%$, $\phi_{mi} = 0.7$, $v_i = 2$~km~s$^{-1}$). The far left hand panel shows the initial distribution of chondrules (light grey) and matrix (dark grey). The colour-scale for the right panels of each of the remaining images denotes the instantaneous stress; the colour-scale for the left panels denotes the temperature. Upon impact, shockwaves are generated at the flyer-sample interface and propagate both down into the sample and up into the flyer plate, compacting the matrix. In this example, after approximately 14.8 $\mu$s the shockwave in the sample has reached the sample/buffer plate interface; by this time the shock in the flyer plate has reflected off the rear of the flyer plate as a release wave that propagates back through the flyer, sample and buffer plates. By 32 $\mu$s the release wave has left the computational domain and the post-shock state of the sample may be recorded. The variation in peak pressure, peak and post-shock temperature experienced by both the chondrules and matrix was recorded for subsequent analysis as was the reduction in porosity in the matrix. See text for details.}
    \label{Fig:ShockPropagation}
\end{figure*}

To generate a shock or impact-induced compaction wave in the simulated chondritic precursor material, numerical planar impact experiments were performed in which a flyer plate impacted a target, comprising a sample sandwiched between a cover plate above and a buffer plate below. The flyer, cover, sample and buffer plates were all composed of the same bimodal mixture of nonporous chondrule disks, surrounded by a high porosity matrix, to eliminate any unwanted wave reflections. The presence of a cover plate allowed the planar shock wave to achieve a steady form before passing through the sample and then the adjacent buffer plate. 

\begin{figure}[!tbp]
    \begin{center}
        \includegraphics[width=3.4in]{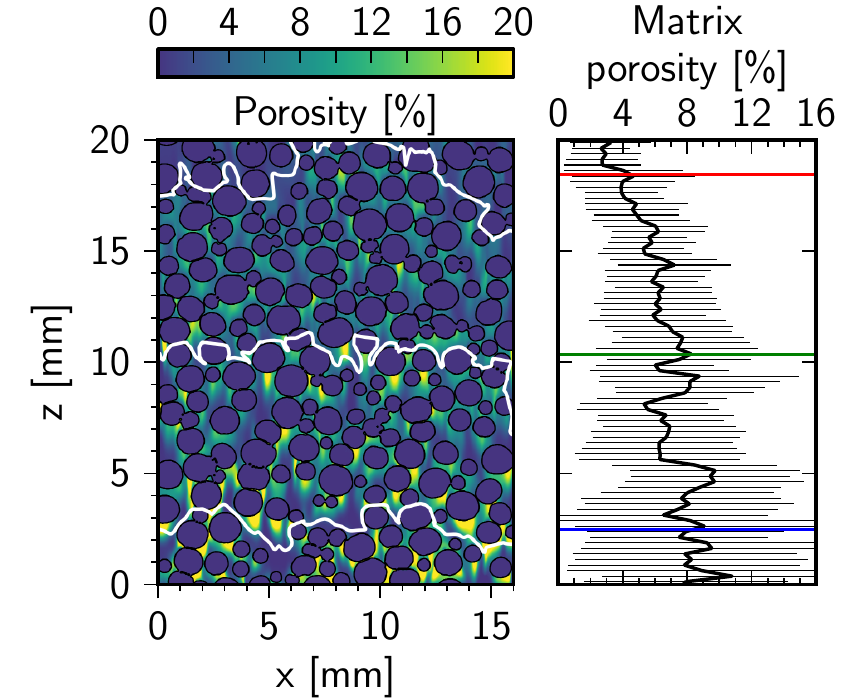}
    \end{center}
    \caption{Porosity field in the cover and sample plates after shock compaction. White lines on the left denote the boundary between the flyer and cover plates (top), the cover and sample (middle) and the sample and buffer plate (bottom); the average position of these boundaries are shown by the red, green and blue lines on the right, respectively. The panel on the right shows the average porosity in the matrix along horizontal strips, 5 computational cells wide (error bars show 1-$\sigma$ variations). In the cover plate, there is a gradient in the porosity, rising from 0.04 at the flyer plate boundary to 0.08 at the sample plate boundary. In the sample, there is no systematic porosity gradient, just local variations, showing that the shock wave has reached a steady state.}
    \label{Fig:CoverPlate}
\end{figure}

Figure~\ref{Fig:ShockPropagation} shows the propagation of a shock wave in a typical simulation from this study. The first panel (on the left) shows the initial make up of the numerical experiment: the flyer plate at the top of the mesh (extending out the top of the image shown), which impacts the cover plate. This generates a shock wave that travels into both the flyer and cover plates (second panel). At early times ($t\approx3$~$\mu$s in Fig.~\ref{Fig:ShockPropagation}), the shock front is unsteady owing to grain-scale reverberations, which tend to diffuse the shock front thickness over a distance related to the grain diameter. In the simulations presented here, the shock front thickness was typically $\sim$2 chondrule diameters, consistent with front thicknesses determined by mesoscale simulations of granular material compaction \citep{Benson:97}. A consequence of this initial increase in shock front thickness is that shock compaction (and hence peak pressure) is greatest at the impact plane and decays with distance until a steady shock wave is achieved. This is evidenced by a gradient in porosity in the cover (and flyer) plate near the interface between them (Fig.~\ref{Fig:CoverPlate}).

A thickness of the cover plate of several chondrule diameters was chosen to ensure that the shock wave was steady when it entered the sample. Hence, by the time the shock wave propagated into the sample plate ($t=7$~$\mu$s in Fig.~\ref{Fig:ShockPropagation}), it had achieved a constant shock front thickness and rise time, which resulted in no gradient in compaction within the sample (see Figure~\ref{Fig:CoverPlate}).  

Despite achieving a steady wave amplitude and shock front thickness, the propagating shock wave exhibited resonant oscillations around the steady wave amplitude, caused by the mesoscale structure of the chondrule-matrix mixture. Such oscillations have been noted in previous models of shock compaction of porous granular materials \citep[e.g.][]{Trott:07} and porous rocks \citep{Gueldemeister:13}. They have also been observed in laboratory experiments of layered composites of `hard' and `soft' materials \citep[e.g.][]{Zhuang:03}. These experiments revealed that the magnitude and duration of the oscillations depends on the impedance mismatch between the components in the system \citep{Zhuang:03}, which is very large for the chondrule -- matrix system studied here. It is these violent oscillations that result in heterogeneous heating of the matrix within the sample \citep{Bland:14}.

The dimensions of the flyer plate were designed such that the release wave generated when the shock wave reflected off the rear end of the flyer plate did not reach the sample plate material until the shock front had already propagated through the entire sample region ($t=14.8$~$\mu$s in Fig.~\ref{Fig:ShockPropagation}). The simulation time extended until the sample was released from high pressure by a release wave from the rear of the flyer plate ($t=32$~$\mu$s inFigure~\ref{Fig:ShockPropagation}).

\subsection{Diagnosis of Bulk and Component Response}
In all simulations, Lagrangian tracer particles monitored the response to shock wave passage of computational-cell-sized parcels of material. This complete record of material history allowed us to construct the response to shock of the individual components (chondrule and matrix) and the bulk material, as well as to document spatial variations in response within the chondrules and insterstitial matrix. The tracer particles were placed throughout the computational mesh (one tracer per cell) at the beginning of the simulation. These particles then tracked the movement of that volume of material throughout the simulation. The advantage of using Lagrangian tracers here, rather than the Eulerian cell-based quantities, it that they can record both the pressure, temperature and porosity at each timestep, and the peak temperature and pressure that each tracer experienced during the entire simulation. Using the cell-based quantities alone would not allow us to track the material's history, and thus the peak quantities.  

In post-processing, each bulk property (final and peak temperature; final and peak pressure; porosity) in the sample was calculated as the volume-weighted average of that variable for all the tracers in the sample plate behind the shock wave (using the tracer volume at the final time). We used a tracer's location to diagnose whether it had been shocked or not: once the tracer started moving (i.e. its displacement was greater than a threshold value), it was determined to be behind the shock wave.

The tracer records also allowed the properties of the individual components (matrix and chondrules) to be calculated, as each tracer represented only one of the two components (see Tables~\ref{Tab:Bulk} and \ref{Tab:Final}). Values of post-shock temperature and porosity were recorded just after the release wave had passed through the sample mixture (e.g., $\sim$~32~$\mu$s in Figure~\ref{Fig:ShockPropagation}). Hence, these temperatures represent the temperature of the material immediately after the passage of the shock and release waves, but prior to the (likely rapid) equilibration of heat between the matrix and chondrules, which is not accounted for in our models and is discussed later (Section~\ref{Sect:EquilibrationTime}).

\begin{figure*}[tb]
    \begin{center}
        \includegraphics[width=\textwidth]{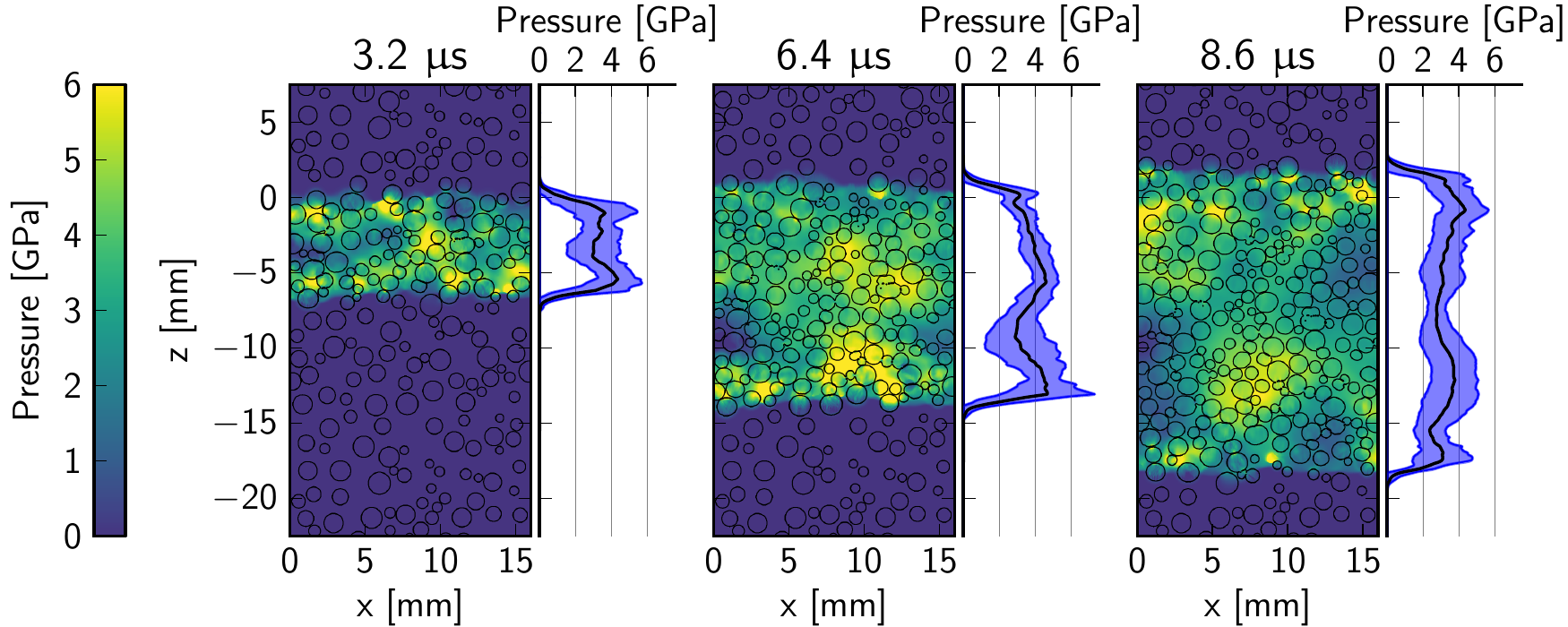}
    \end{center}
    \caption{The pressure within the shock wave at three timesteps (3.2~$\mu$s, 6.4~$\mu$s and 8.6~$\mu$s), for the reference calculation shown in Figure~\ref{Fig:ShockPropagation} ($A_{mi} = 70\%$, $\phi_{mi} = 0.7$, $v_i = 2$~km~s$^{-1}$). The line graphs show the pressure averaged along each row of cells in the $z$ direction. The blue shaded region shows the 1-$\sigma$ variation in pressure along each row of cells. At any given time, the pressure in the shock wave can vary by several GPa, on lengthscales of a few millimetres.}
    \label{Fig:BulkCalc}
\end{figure*}

A consequence of the heterogeneity of the sample is that the shock wave is also heterogenous, and thus the peak shock pressure recorded by any tracer throughout the duration of shockwave passage can be substantially higher than the instantaneous bulk shock pressure at any time. For example, in the simulation shown in Figure~\ref{Fig:ShockPropagation}, the average bulk shock pressure in the shockwave was approximately 3.0~GPa and yet the mean peak shock pressure experienced by chondrule and matrix material in the sample was 6.5~GPa and 8.4~GPa respectively. This can be seen in Figure~\ref{Fig:BulkCalc}, which shows the variation in pressure within the shock wave at three different timesteps. The line graphs show the average pressure along each row of computational cells, with the 1-$\sigma$ variation in pressure denoted by the blue shaded region. Within the shock wave, these average pressures can vary by as much as 2 GPa within a space of 5 mm. This behaviour has been observed previously in models of compaction of porous sandstone, where a peak pressure was recorded of up to 4 times greater than the bulk pressure \citep{Gueldemeister:13}. 

\subsubsection{Lagrangian tracer motion}

Due to the large number of mixed-material computational cells in these simulations with many small particles supported in a continuous matrix, it is important to make sure that the Lagrangian tracers used to track material history stay with their respective materials. The standard approach in iSALE is to move tracers through the computational mesh using velocities interpolated from the surrounding nodal velocities. However, it was found that in mixed cells, some tracers ``drifted" into neighbouring materials (e.g. chondrule tracers ended up in a cell full of matrix, etc.). To address this problem, a new method of calculating tracer velocities was devised using material volume fluxes. This is documented fully in Appendix~\ref{Apdx:Tracers}.

Figure~\ref{Fig:Tracers} shows a comparison of a simulation run using the old (velocity) method and the new (material) method. Note that in areas where chondrules (black tracers) have collided, in the velocity method some matrix tracers have drifted into the chondrules, whereas in the material method, all tracers respect the material boundaries. The new material method is used throughout the work presented here.

\begin{figure}[tb]
    \begin{center}
        \includegraphics[width=3.4in]{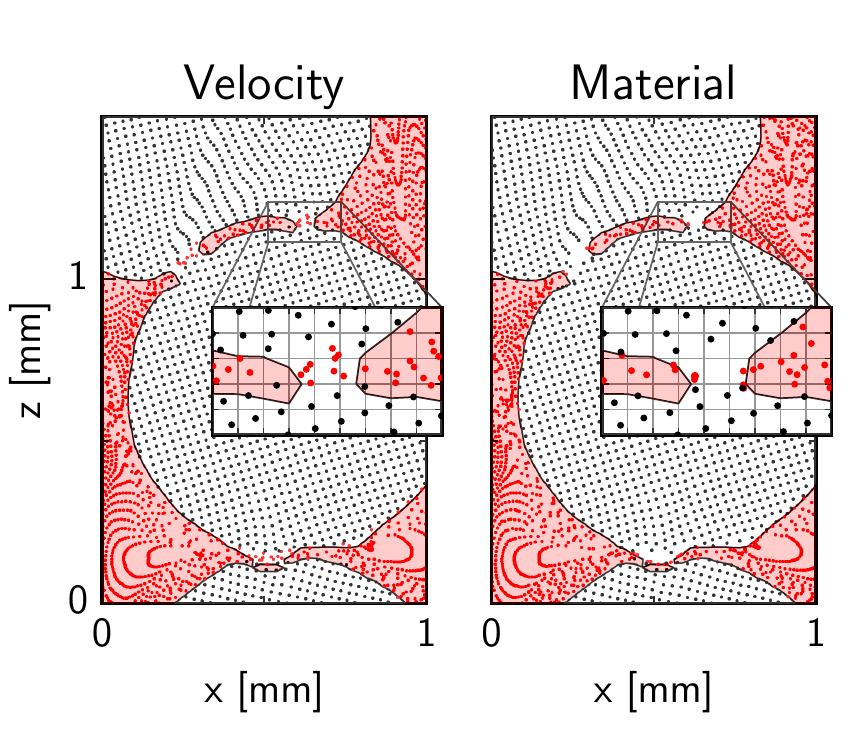}
    \end{center}
    \caption{A comparison of the two methods in iSALE for moving tracers during a timestep. The velocity method (left) allows tracers to drift out of their respective materials, whereas the new material-based method (right) respects material boundaries. Note that where some tracers appear to have crossed the material boundary in the material method, they are in mixed-material cells (i.e. a cell with a material boundary in it) and thus are still attached to their respective material --- this is just a result of how the material boundary contours were drawn while constructing the figure.}
    \label{Fig:Tracers}
\end{figure}

\subsection{Resolution}
\begin{figure}[!t]
    \centering
    \includegraphics[width=3.4in]{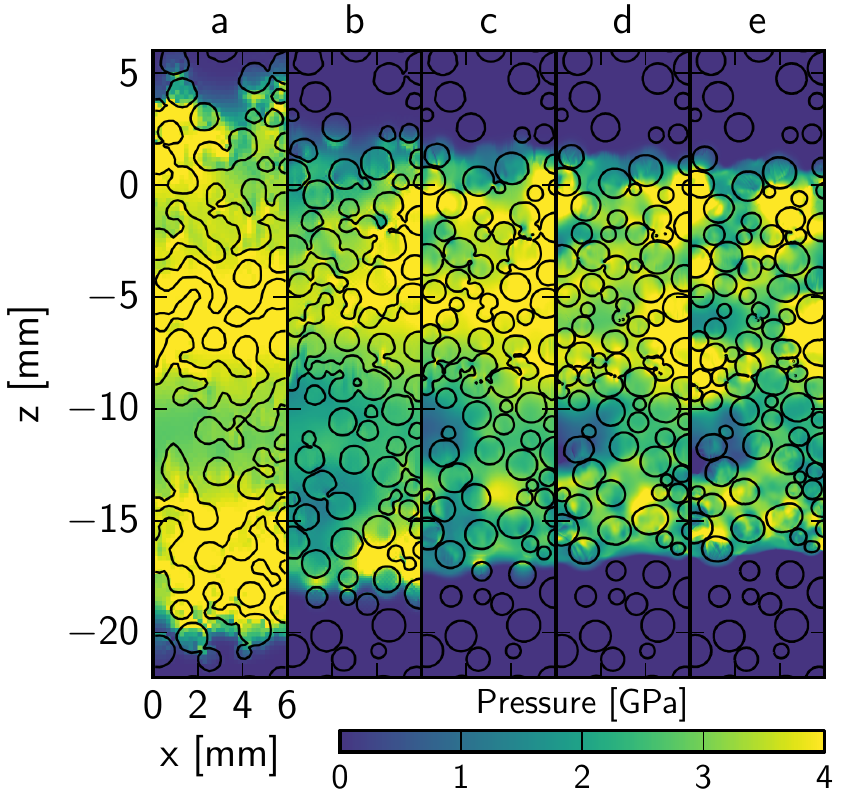}
    \caption{Pressure in a suite of simulations with different resolutions, after 80~$\mu$s of model time. A cell size of (a) 200~\micron\ (2.5 cells per particle radius); (b) 100~\micron\ (5 cppr); (c) 50~\micron\ (10cppr); (d) 25~\micron\ (20 cppr); and (e) 12.5~\micron\ (40 cppr).}
    \label{Fig:ResolutionMeshes}
\end{figure}
\begin{figure}[!t]
    \centering
    \includegraphics[width=3.4in]{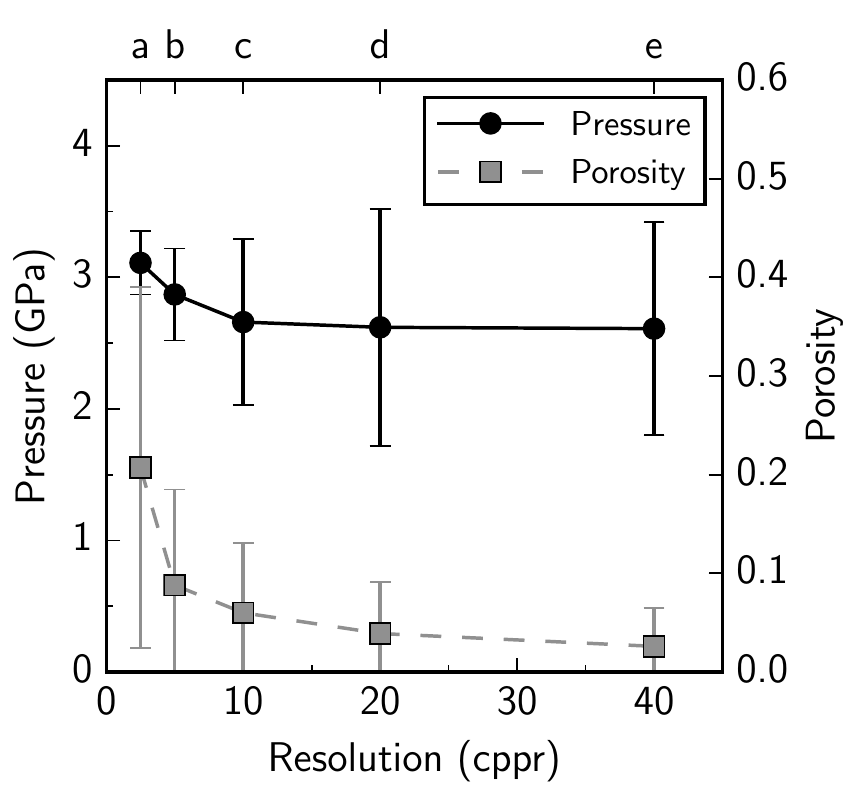}
    \caption{Peak shock pressure and porosity in the bulk sample as a function of model resolution. Black circles show the volume-weighted mean of peak pressure experienced by all tracers in the sample; grey squares show the volume-weighted mean of porosity. Error bars show the 1-$\sigma$ variation in peak pressure or porosity among the same tracers. Labels on top axis correspond to the five simulations shown in Figure~\ref{Fig:ResolutionMeshes}.}
    \label{Fig:Resolution}
\end{figure}

Simulations were run at various resolutions, ranging from a mean value of 40 cells per particle radius (cppr) to 2.5 cppr (cell sizes of 12.5~$\mu$m to 200~$\mu$m), with an identical particle distribution and number of tracer particles in each simulation (Figure~\ref{Fig:ResolutionMeshes}). The mean pressure converges at around 10 cppr (Figure~\ref{Fig:Resolution}), but differences in the shock speed and structure within the shock are still discernible between 10 and 20 cppr (Figure~\ref{Fig:ResolutionMeshes}d \& \ref{Fig:ResolutionMeshes}e). The mean porosity converges at around 20 cppr.

The error bars on Figure~\ref{Fig:Resolution} show the standard deviation in pressure in that simulation (they are not a measure of model error). For simulations with a resolution of at least 5 cppr, the mean pressures all fell within the 1-$\sigma$ variation in pressure for all other resolutions; the same is true for porosity for simulations with 10cppr or above (a similar trend is seen for temperature). As a compromise between computational expense and time, a mean particle radius of 20 cppr was used for the remainder of this study.

\subsection{Shock Duration}\label{Sect:ShockDuration}
\begin{figure}[t]
    \centering
    \includegraphics[width=3.4in]{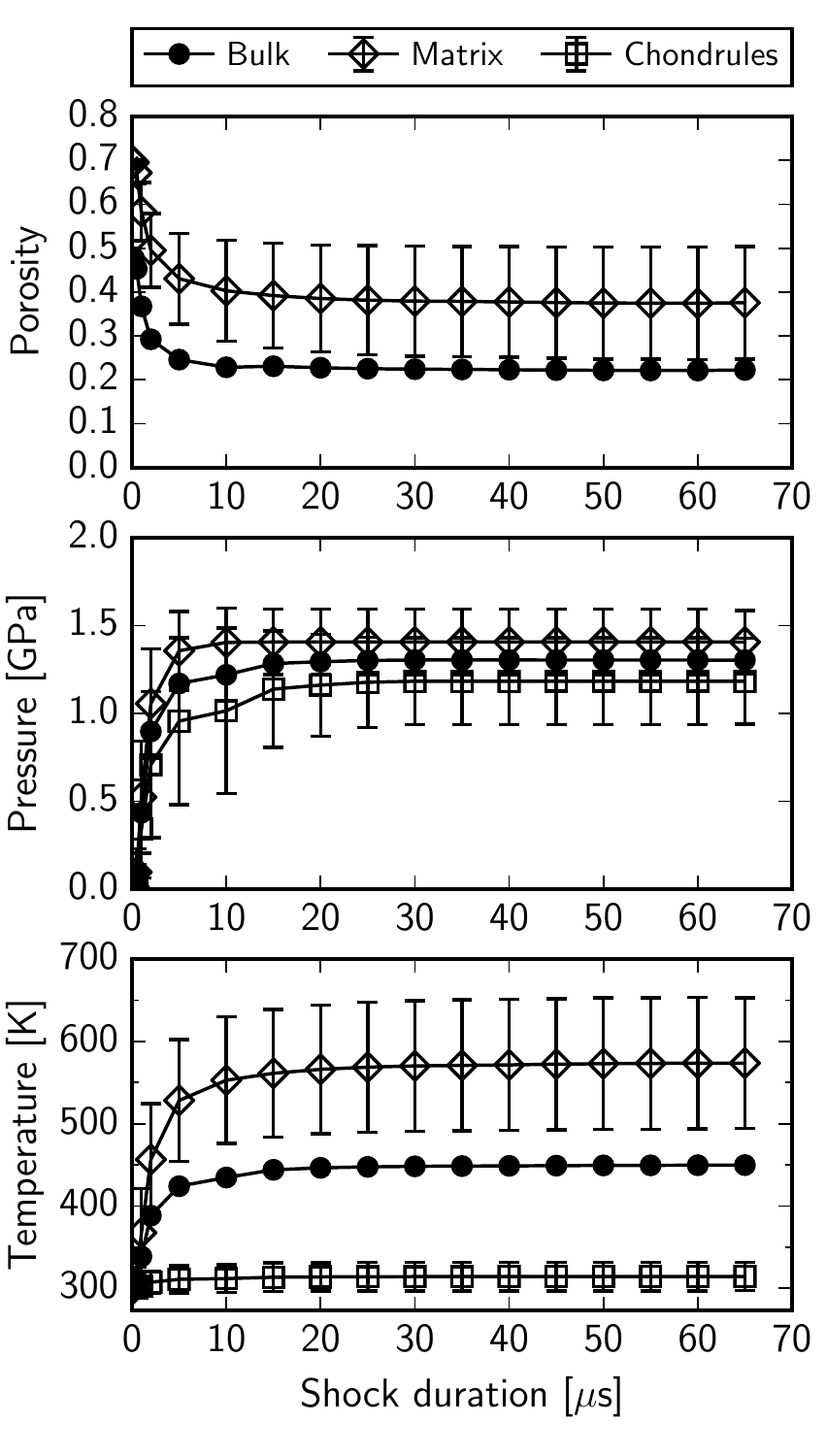}
    \caption{Final porosity, peak pressure and final temperature as a function of shock duration.}
    \label{Fig:ShockDuration}
\end{figure}

The duration of the shock pulse in the simulations presented below (Section~\ref{Sect:Results} and Tables~\ref{Tab:Bulk} and \ref{Tab:Final}) (i.e. the time from shock to release) ranges from 30 to 80 $\mu$s (90\% of the simulations presented here have a shock duration of at least 40 $\mu$s). It should be noted that this shock duration is significantly lower than the duration expected on a meteorite parent body during an impact event, but longer than the shock durations that are possible in typical gas-gun experiments. To investigate the shock duration required in order to reach a steady state (and thus allow any simulation results to be applied to the larger planetesimal scale), a simulation was run of two colliding 120 mm long impactors, with a similar composition to that shown in Figure~\ref{Fig:ShockPropagation} and a mutual impact velocity of 1~\kms. For all tracers, the time of the shock wave arrival and the time of release were recorded; from these measurements, the shock duration experienced by each tracer was calculated. The mean and standard deviation of the peak pressure, final porosity and final temperature among those tracers in the shock were recorded as a function of shock duration. Figure~\ref{Fig:ShockDuration} shows that for shock durations of 20--30 $\mu$s or longer, the porosity, pressure and temperature measured in the simulation has reached a steady state. This suggests that the results presented here should apply to impacts over a large range of scales (with longer shock durations) including planetesimal-scale collisions.

\section{RESULTS}\label{Sect:Results}
\begin{table*}[!t]
\begin{center}

\caption{Bulk and compaction properties}

\begin{tabular}{llcccp{0.1ex}ccccccc}
\hline\hline
& & & & & & & & \\[-8pt]
\multicolumn{3}{l}{Initial conditions} & & & & \multicolumn{6}{l}{Compaction and bulk properties}\\
\\
\cline{1-5} \cline{7-12}
& & & & & & & & \\[-8pt]
Matrix & $V_i$ & $A_{mi}$\tablenotemark{a} & $\phi_{mi}$\tablenotemark{b} & $\phi_{bi}$\tablenotemark{c} & & $A_{mf}$\tablenotemark{d} & $\phi_{mf}$\tablenotemark{e} & $\phi_{bf}$\tablenotemark{f} & $P_{sh}$\tablenotemark{g} & $T_{bf}$\tablenotemark{h} & $T_{bp}$\tablenotemark{i} \\
material & (\kms)  &   (\%)          &         &  &            & (\%)     &        &     & (GPa)    & (K)      & (K)\\[2pt]
\hline
& & & & & & & & \\[-4pt]
Dunite     & 0.75 & 82 & 0.7 & 0.58 && 74 & 0.51 $\pm$ 0.03 & 0.38 & 0.56 & 370  & 376   \\
           & 1    & 81 & 0.7 & 0.56 && 68 & 0.40 $\pm$ 0.06 & 0.27 & 0.83 & 419  & 423   \\
           & 1.25 & 80 & 0.7 & 0.56 && 64 & 0.29 $\pm$ 0.09 & 0.19 & 1.03 & 470  & 475   \\
           & 1.5  & 81 & 0.7 & 0.57 && 62 & 0.17 $\pm$ 0.09 & 0.11 & 1.33 & 544  & 551   \\
           & 2    & 81 & 0.7 & 0.57 && 59 & 0.06 $\pm$ 0.05 & 0.04 & 2.17 & 704  & 717   \\
           & 0.75 & 70 & 0.7 & 0.49 && 58 & 0.48 $\pm$ 0.04 & 0.28 & 0.64 & 371  & 377   \\
           & 1    & 69 & 0.6 & 0.42 && 52 & 0.17 $\pm$ 0.08 & 0.09 & 1.14 & 408  & 414   \\
           & 1    & 72 & 0.7 & 0.50 && 54 & 0.33 $\pm$ 0.08 & 0.18 & 0.98 & 425  & 430   \\
           & 1    & 72 & 0.8 & 0.58 && 54 & 0.54 $\pm$ 0.07 & 0.29 & 0.70 & 451  & 454   \\
           & 1.5  & 71 & 0.7 & 0.49 && 45 & 0.10 $\pm$ 0.08 & 0.05 & 1.71 & 539  & 548   \\
           & 2    & 72 & 0.6 & 0.43 && 51 & 0.04 $\pm$ 0.03 & 0.02 & 3.98 & 674  & 696   \\
           & 2    & 71 & 0.7 & 0.50 && 45 & 0.07 $\pm$ 0.04 & 0.03 & 3.02 & 701  & 712   \\
           & 2    & 69 & 0.8 & 0.55 && 34 & 0.09 $\pm$ 0.09 & 0.03 & 1.92 & 715  & 726   \\
           & 2.5  & 70 & 0.7 & 0.49 && 43 & 0.05 $\pm$ 0.03 & 0.02 & 5.16 & 897  & 935   \\
           & 3    & 71 & 0.7 & 0.50 && 45 & 0.04 $\pm$ 0.03 & 0.02 & 7.62 & 1120 & 1200  \\
           & 1    & 62 & 0.7 & 0.44 && 43 & 0.31 $\pm$ 0.08 & 0.14 & 1.12 & 422  & 429   \\
           & 1.5  & 62 & 0.7 & 0.43 && 35 & 0.09 $\pm$ 0.06 & 0.03 & 2.36 & 536  & 550   \\
           & 2    & 63 & 0.7 & 0.44 && 36 & 0.07 $\pm$ 0.03 & 0.03 & 4.38 & 702  & 725   \\
           & 1    & 52 & 0.7 & 0.36 && 31 & 0.28 $\pm$ 0.09 & 0.09 & 1.39 & 420  & 428   \\
           & 1.5  & 50 & 0.7 & 0.35 && 25 & 0.09 $\pm$ 0.05 & 0.02 & 3.09 & 536  & 548   \\
           & 2    & 50 & 0.7 & 0.35 && 26 & 0.09 $\pm$ 0.04 & 0.02 & 5.76 & 680  & 710   \\
           & 1    & 42 & 0.7 & 0.29 && 23 & 0.25 $\pm$ 0.08 & 0.06 & 1.73 & 417  & 426   \\
           & 1.5  & 41 & 0.7 & 0.29 && 20 & 0.12 $\pm$ 0.06 & 0.02 & 3.84 & 531  & 548   \\
           & 2    & 42 & 0.7 & 0.29 && 20 & 0.11 $\pm$ 0.05 & 0.02 & 7.25 & 681  & 718   \\
           & 0.75 & 33 & 0.7 & 0.23 && 20 & 0.38 $\pm$ 0.08 & 0.08 & 1.34 & 363  & 376   \\
           & 1    & 33 & 0.7 & 0.23 && 16 & 0.23 $\pm$ 0.09 & 0.04 & 2.16 & 410  & 422   \\
           & 1.5  & 33 & 0.7 & 0.23 && 15 & 0.14 $\pm$ 0.07 & 0.02 & 5.12 & 530  & 552   \\
           & 2    & 33 & 0.7 & 0.23 && 15 & 0.13 $\pm$ 0.06 & 0.02 & 9.40 & 689  & 705   \\
           & 2.5  & 33 & 0.7 & 0.23 && 15 & 0.13 $\pm$ 0.07 & 0.02 & 14.1 & 874  & 894   \\
           & 3    & 34 & 0.7 & 0.24 && 15 & 0.13 $\pm$ 0.07 & 0.02 & 18.9 & 1130 & 1200  \\[8pt]
Serpentine & 1    & 69 & 0.6 & 0.41 && 49 & 0.05 $\pm$ 0.07 & 0.03 & 0.87 & 397  & 400   \\
           & 2    & 69 & 0.6 & 0.41 && 48 & 0.00 $\pm$ 0.01 & 0.00 & 3.69 & 595  & 609   \\
           & 3    & 71 & 0.6 & 0.43 && 51 & 0.00 $\pm$ 0.00 & 0.00 & 8.20 & 901  & 955   \\
           & 1    & 70 & 0.7 & 0.49 && 47 & 0.20 $\pm$ 0.11 & 0.10 & 0.72 & 405  & 406   \\
           & 2    & 70 & 0.7 & 0.49 && 42 & 0.01 $\pm$ 0.02 & 0.00 & 2.60 & 615  & 626   \\
           & 3    & 70 & 0.7 & 0.49 && 42 & 0.00 $\pm$ 0.00 & 0.00 & 6.47 & 947  & 977   \\[4pt]
\hline
\end{tabular}
\tablenotetext{1}{Initial matrix abundance}
\tablenotetext{2}{Initial matrix porosity}
\tablenotetext{3}{Initial bulk porosity}
\tablenotetext{4}{Final matrix abundance}
\tablenotetext{5}{Final matrix porosity (with 1-$\sigma$ variations)}
\tablenotetext{6}{Final bulk porosity}
\tablenotetext{7}{Peak shock pressure}
\tablenotetext{8}{Final bulk temperature}
\tablenotetext{9}{Peak bulk temperature}
\label{Tab:Bulk}
\end{center}
\end{table*}

\begin{table*}[!t]
\begin{center}

\caption{Matrix and chondrule properties}

\begin{tabular}{llccp{0.1ex}cccccc}
\hline\hline
& & & & & & & & \\[-8pt]
\multicolumn{3}{l}{Initial conditions} & & & \multicolumn{6}{l}{Matrix and chondrule properties}\\
\\
\cline{1-4} \cline{6-11}
& & & & & & & & \\[-8pt]
Matrix & $V_i$ & $A_{mi}$\tablenotemark{a} & $\phi_{mi}$\tablenotemark{b} & & $P_{mp}$\tablenotemark{c} & $T_{mf}$\tablenotemark{d} & $T_{mp}$\tablenotemark{e} & $P_{cp}$\tablenotemark{f} & $T_{cf}$\tablenotemark{g} & $T_{cp}$\tablenotemark{h}\\
material & (\kms)  &  (\%)           &    &  & (GPa)    & (K)      & (K)  & (GPa)    & (K)      & (K)\\[4pt]
\hline
& & & & & & & & \\[-4pt]
Dunite     & 0.75 & 82 & 0.7 && 0.65 $\pm$ 0.04 & 396  $\pm$ 18.1 & 405  $\pm$ 19.3 & 0.58 $\pm$ 0.08 & 300 $\pm$ 0.41 & 303 $\pm$ 1.22 \\
           & 1    & 81 & 0.7 && 1.00 $\pm$ 0.10 & 478  $\pm$ 32.0 & 487  $\pm$ 34.2 & 0.88 $\pm$ 0.13 & 301 $\pm$ 1.05 & 305 $\pm$ 4.30 \\
           & 1.25 & 80 & 0.7 && 1.47 $\pm$ 0.24 & 572  $\pm$ 58.2 & 588  $\pm$ 63.9 & 1.28 $\pm$ 0.29 & 302 $\pm$ 4.12 & 309 $\pm$ 12.2 \\
           & 1.5  & 81 & 0.7 && 2.01 $\pm$ 0.34 & 701  $\pm$ 62.3 & 723  $\pm$ 72.8 & 1.76 $\pm$ 0.42 & 303 $\pm$ 7.64 & 313 $\pm$ 19.9 \\
           & 2    & 81 & 0.7 && 5.54 $\pm$ 1.77 & 973  $\pm$ 76.3 & 1040 $\pm$ 111  & 4.27 $\pm$ 0.96 & 324 $\pm$ 26.3 & 347 $\pm$ 49.0 \\
           & 0.75 & 70 & 0.7 && 0.80 $\pm$ 0.10 & 424  $\pm$ 29.9 & 436  $\pm$ 31.6 & 0.71 $\pm$ 0.19 & 301 $\pm$ 2.28 & 304 $\pm$ 4.19 \\
           & 1    & 69 & 0.6 && 1.70 $\pm$ 0.29 & 507  $\pm$ 38.3 & 521  $\pm$ 44.2 & 1.45 $\pm$ 0.37 & 303 $\pm$ 7.46 & 311 $\pm$ 18.8 \\
           & 1    & 72 & 0.7 && 1.28 $\pm$ 0.17 & 533  $\pm$ 43.3 & 545  $\pm$ 47.1 & 1.11 $\pm$ 0.27 & 302 $\pm$ 4.78 & 308 $\pm$ 10.9 \\
           & 1    & 72 & 0.8 && 0.96 $\pm$ 0.15 & 589  $\pm$ 71.8 & 603  $\pm$ 76.9 & 0.85 $\pm$ 0.23 & 302 $\pm$ 1.93 & 306 $\pm$ 6.15 \\
           & 1.5  & 71 & 0.7 && 3.11 $\pm$ 0.82 & 825  $\pm$ 89.7 & 862  $\pm$ 105  & 2.68 $\pm$ 0.72 & 310 $\pm$ 18.6 & 326 $\pm$ 39.6 \\
           & 2    & 72 & 0.6 && 10.2 $\pm$ 2.51 & 934  $\pm$ 108  & 1050 $\pm$ 236  & 8.71 $\pm$ 2.36 & 400 $\pm$ 52.8 & 438 $\pm$ 76.2 \\
           & 2    & 71 & 0.7 && 8.36 $\pm$ 2.61 & 1110 $\pm$ 110  & 1220 $\pm$ 193  & 6.53 $\pm$ 1.79 & 367 $\pm$ 45.2 & 396 $\pm$ 71.0 \\
           & 2    & 69 & 0.8 && 4.92 $\pm$ 1.56 & 1430 $\pm$ 167  & 1520 $\pm$ 220  & 4.13 $\pm$ 1.18 & 353 $\pm$ 41.3 & 380 $\pm$ 77.0 \\
           & 2.5  & 70 & 0.7 && 13.1 $\pm$ 3.29 & 1450 $\pm$ 185  & 1720 $\pm$ 448  & 10.8 $\pm$ 2.77 & 476 $\pm$ 77.0 & 528 $\pm$ 109 \\
           & 3    & 71 & 0.7 && 16.9 $\pm$ 3.93 & 1790 $\pm$ 270  & 2240 $\pm$ 651  & 14.8 $\pm$ 3.39 & 565 $\pm$ 98.3 & 642 $\pm$ 129 \\
           & 1    & 62 & 0.7 && 1.63 $\pm$ 0.31 & 587  $\pm$ 68.3 & 607  $\pm$ 73.7 & 1.46 $\pm$ 0.43 & 304 $\pm$ 8.85 & 313 $\pm$ 20.8 \\
           & 1.5  & 62 & 0.7 && 4.43 $\pm$ 1.28 & 930  $\pm$ 113  & 989  $\pm$ 141  & 3.77 $\pm$ 1.02 & 326 $\pm$ 33.3 & 351 $\pm$ 65.5 \\
           & 2    & 63 & 0.7 && 9.64 $\pm$ 2.23 & 1245 $\pm$ 149  & 1400 $\pm$ 257  & 7.79 $\pm$ 1.53 & 399 $\pm$ 51.5 & 439 $\pm$ 87.0 \\
           & 1    & 52 & 0.7 && 1.99 $\pm$ 0.52 & 661  $\pm$ 100  & 684  $\pm$ 111  & 1.90 $\pm$ 0.64 & 313 $\pm$ 22.3 & 327 $\pm$ 43.3 \\
           & 1.5  & 50 & 0.7 && 5.59 $\pm$ 1.34 & 1063 $\pm$ 151  & 1140 $\pm$ 181  & 4.77 $\pm$ 1.03 & 355 $\pm$ 45.3 & 383 $\pm$ 82.2 \\
           & 2    & 50 & 0.7 && 11.2 $\pm$ 2.12 & 1412 $\pm$ 213  & 1610 $\pm$ 326  & 9.43 $\pm$ 1.59 & 426 $\pm$ 64.1 & 473 $\pm$ 106 \\
           & 1    & 42 & 0.7 && 2.62 $\pm$ 0.84 & 754  $\pm$ 128  & 789  $\pm$ 142  & 2.51 $\pm$ 0.82 & 318 $\pm$ 25.5 & 335 $\pm$ 48.3 \\
           & 1.5  & 41 & 0.7 && 6.95 $\pm$ 1.63 & 1200 $\pm$ 201  & 1310 $\pm$ 237  & 6.06 $\pm$ 1.20 & 367 $\pm$ 55.0 & 405 $\pm$ 95.2 \\
           & 2    & 42 & 0.7 && 12.4 $\pm$ 2.16 & 1590 $\pm$ 260  & 1830 $\pm$ 366  & 10.8 $\pm$ 1.59 & 451 $\pm$ 73.5 & 510 $\pm$ 119 \\
           & 0.75 & 33 & 0.7 && 1.87 $\pm$ 0.77 & 600  $\pm$ 123  & 640  $\pm$ 138  & 1.96 $\pm$ 0.74 & 307 $\pm$ 24.4 & 324 $\pm$ 33.1 \\
           & 1    & 33 & 0.7 && 3.22 $\pm$ 1.22 & 842  $\pm$ 161  & 884  $\pm$ 183  & 3.23 $\pm$ 0.98 & 326 $\pm$ 32.3 & 349 $\pm$ 53.2 \\
           & 1.5  & 33 & 0.7 && 7.75 $\pm$ 1.71 & 1330 $\pm$ 235  & 1460 $\pm$ 278  & 7.18 $\pm$ 1.33 & 383 $\pm$ 57.1 & 428 $\pm$ 96.7 \\
           & 2    & 33 & 0.7 && 13.9 $\pm$ 2.24 & 1810 $\pm$ 320  & 2130 $\pm$ 424  & 12.4 $\pm$ 1.58 & 448 $\pm$ 73.6 & 519 $\pm$ 116 \\
           & 2.5  & 33 & 0.7 && 21.1 $\pm$ 3.08 & 2410 $\pm$ 409  & 2990 $\pm$ 621  & 18.8 $\pm$ 2.15 & 546 $\pm$ 81.5 & 625 $\pm$ 109 \\
           & 3    & 34 & 0.7 && 27.8 $\pm$ 3.91 & 3020 $\pm$ 436  & 3880 $\pm$ 758  & 24.5 $\pm$ 2.77 & 642 $\pm$ 86.9 & 741 $\pm$ 106 \\[8pt]
Serpentine & 1    & 69 & 0.6 && 1.45 $\pm$ 0.43 & 496  $\pm$ 40.3 & 505  $\pm$ 43.6 & 1.26 $\pm$ 0.47 & 305 $\pm$ 9.90 & 314 $\pm$ 24.3 \\
           & 2    & 69 & 0.6 && 8.81 $\pm$ 1.71 & 814  $\pm$ 92.2 & 904  $\pm$ 162  & 7.49 $\pm$ 1.51 & 396 $\pm$ 46.8 & 431 $\pm$ 73.6 \\
           & 3    & 71 & 0.6 && 17.0 $\pm$ 3.01 & 1240 $\pm$ 153  & 1500 $\pm$ 342  & 14.7 $\pm$ 2.98 & 558 $\pm$ 83.4 & 628 $\pm$ 116  \\
           & 1    & 70 & 0.7 && 1.01 $\pm$ 0.22 & 527  $\pm$ 50.2 & 534  $\pm$ 53.8 & 0.92 $\pm$ 0.33 & 302 $\pm$ 3.99 & 308 $\pm$ 11.2 \\
           & 2    & 70 & 0.7 && 7.68 $\pm$ 1.96 & 941  $\pm$ 99.9 & 1020 $\pm$ 168  & 6.14 $\pm$ 1.63 & 379 $\pm$ 47.1 & 410 $\pm$ 69.9 \\
           & 3    & 70 & 0.7 && 15.4 $\pm$ 3.01 & 1440 $\pm$ 165  & 1690 $\pm$ 358  & 13.2 $\pm$ 2.72 & 590 $\pm$ 111  & 649 $\pm$ 153  \\[4pt]
\hline
\end{tabular}
\tablenotetext{1}{Initial matrix abundance}
\tablenotetext{2}{Initial matrix porosity}
\tablenotetext{3}{Peak shock pressure in the matrix (with 1-$\sigma$ variations)}
\tablenotetext{4}{Final matrix temperature (with 1-$\sigma$ variations)}
\tablenotetext{5}{Peak matrix temperature (with 1-$\sigma$ variations)}
\tablenotetext{6}{Peak shock pressure in the chondrule (with 1-$\sigma$ variations)}
\tablenotetext{7}{Final chondrule temperature (with 1-$\sigma$ variations)}
\tablenotetext{8}{Peak chondrule temperature (with 1-$\sigma$ variations)}
\label{Tab:Final}
\end{center}
\end{table*}

The response of the bimodal mixture of chondrules and matrix to a shock wave was simulated using the techniques described in Section~\ref{Sect:Methods}. In Section~\ref{Sect:BaseModel}, the model shown in Figure~\ref{Fig:ShockPropagation} is documented in detail. In the subsequent sections (\ref{Sect:Velocity}~--~\ref{Sect:Material}), a wide parameter space is explored (impact velocity, matrix fraction, matrix porosity and matrix material) and results compared to those of this reference model (see also Tables~\ref{Tab:Bulk}~and~\ref{Tab:Final}). 

The range of impact velocities we consider here is 0.75 -- 3~km~s$^{-1}$. While the average collision velocity in the main asteroid belt today is $\sim$~5~km~s$^{-1}$, dynamical models of terrestrial planet formation and planetesimal collisional histories \citep{Obrien:06, Obrien:07, Davison:13} show that in the first million years of solar system evolution the mean collision velocity of planetesimals was likely $<$~3~km~s$^{-1}$. The velocities range simulated here was also chosen to give post-compaction matrix abundances and porosities consistent with carbonaceous and ordinary chondrites (see Section~\ref{Sect:MeteoriteGroups}).

\subsection{Mesoscale Response to Shock of Bimodal Mixtures} \label{Sect:BaseModel}

Figure~\ref{Fig:ShockPropagation} depicts the evolution of a simulation we use as a reference in discussing our results. In that simulation the impact velocity $v_i$~=~2~\kms, the matrix fraction $A_{mi}$~=~70\% and the initial matrix porosity $\phi_{mi}$~=~0.7. The left hand panel depicts the initial model setup. A flyer plate impacts a cover plate at $v_i$, sending a shockwave into both the cover plate and the flyer (3~$\mu$s). At 7~$\mu$s, the shockwave enters the sample plate (the region over which the statistical analysis will be performed). At 14.8~$\mu$s, the shockwave reaches the back edge of the sample, and enters a buffer plate (the purpose of which is to prevent the generation of a release wave at this location). Also at this time, a release wave is formed at the top edge of the flyer, which starts to move though the flyer towards the cover and sample. By 32~$\mu$s, the entire sample has been released from high pressure.

\begin{figure}[t]
    \centering
    \includegraphics[width=3.4in]{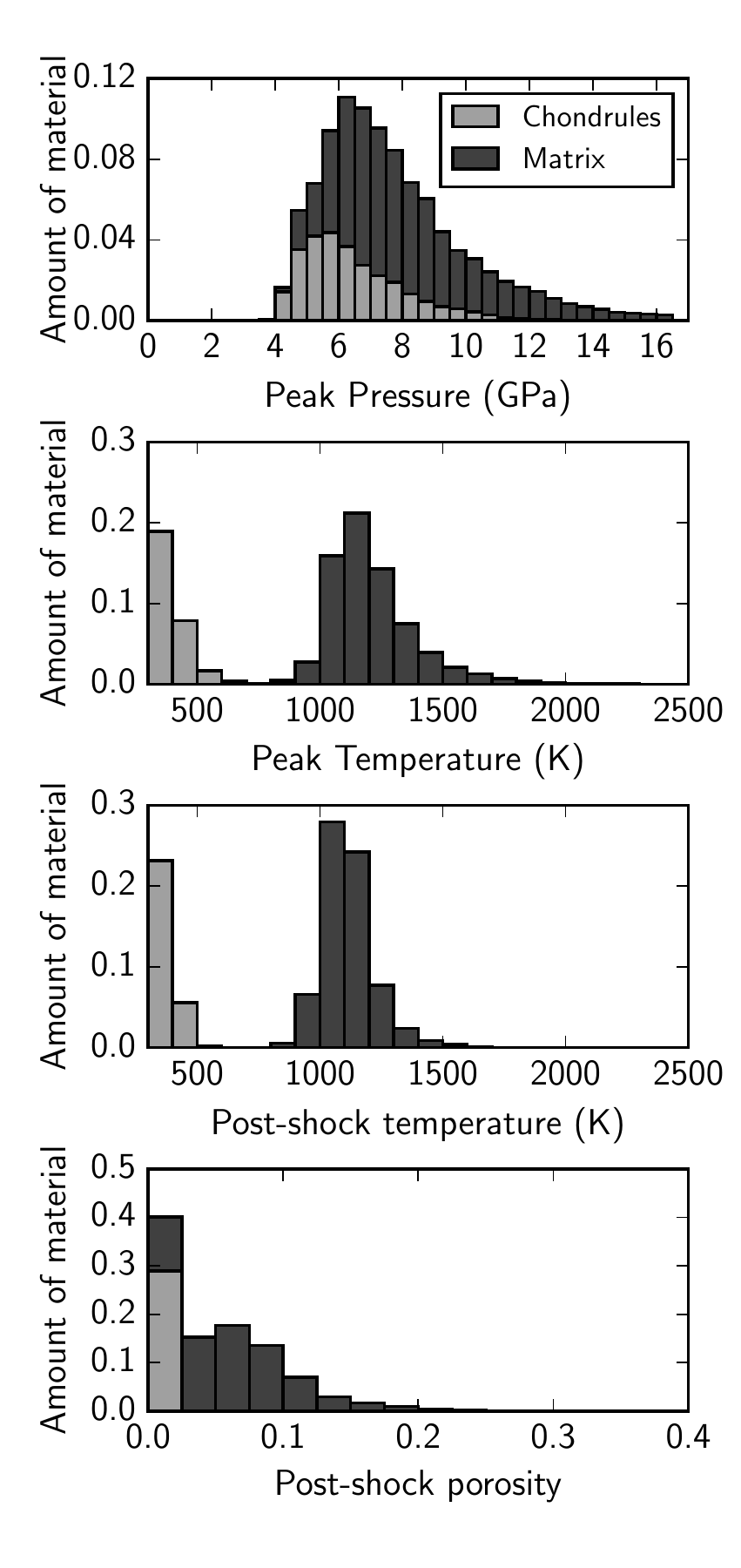}
    \caption{Histograms showing the mesoscopic response observed in the reference simulation (Section~\ref{Sect:BaseModel}) of a bimodal mixture of chondrules and matrix in and impact at 2~\kms into a mixture with $A_{mi}$ = 70\% and $\phi_{mi}$ = 0.7.}
    \label{Fig:Histograms}
\end{figure}

In Figure~\ref{Fig:Histograms} the differences between the chondrules and matrix are highlighted for the reference simulation. The peak pressures in the chondrules range from $\sim$ 4 to 11 GPa, with a mean of 6.5~GPa, and in the matrix from 4 to 16 GPa, with a mean of 8.4~GPa. The pressure is higher in the matrix than in the chondrules, due to the large strength difference between the two components (the longitudinal stress, however, is the same in both components). Both the peak-shock and post-shock temperatures in the chondrules are low (between 300 K and 500 K) compared to temperatures in the matrix (900 K to $>$ 1500 K). The chondrules begin with no porosity, and that is not changed in the shock event; the matrix begins with a porosity of 0.7, and is reduced to a range of 0 -- 0.2 porosity in the shock event, with a mean of 0.07. The bulk porosity, $\phi_{bf}$ is compacted from 0.5 before the impact to 0.03 after the impact, and the abundance of matrix decreases from 70\% to 45\% as a result of that compaction. The location of the residual porosity is shown in the top row of Figure~\ref{Fig:Vel}, and appears as shadow regions on the lee side of the chondrules (the shock direction was down the page), where material is somewhat protected from the shock and compaction is incomplete.

\subsection{Effect of Impact Velocity, $v_{i}$}\label{Sect:Velocity}

\begin{figure*}[t]
    \centering
    \includegraphics[width=7in]{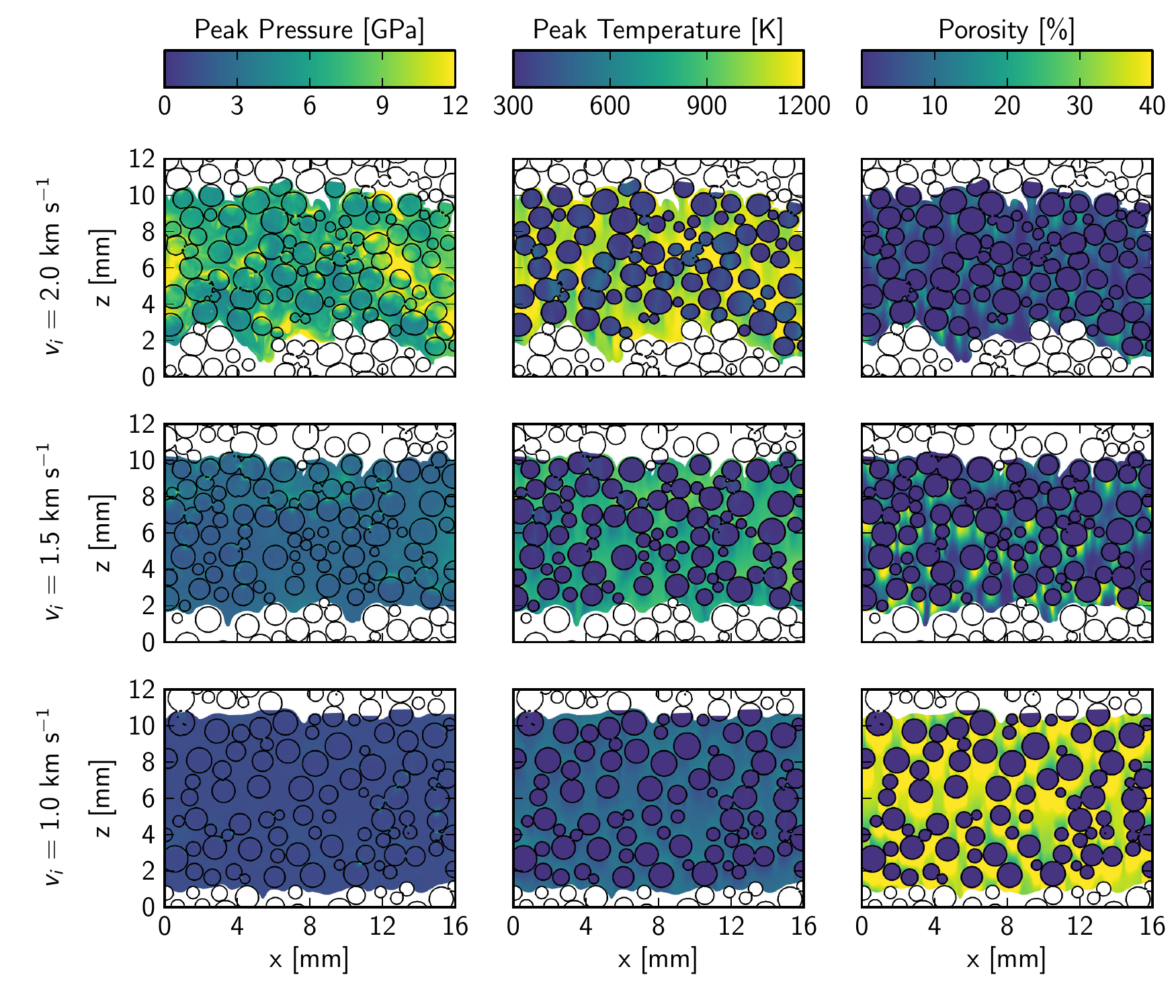}
    \caption{The effect of impact velocity ($v_{i}$) on the peak pressure, final temperature and porosity within the sample zone. Notice also the increase in chondrule deformation with impact speed. In these three simulations, the matrix abundance ($A_{mi}$ = 70\%), matrix porosity ($\phi_{mi}$ = 0.7) and matrix material (dunite) were all kept constant.}
    \label{Fig:Vel}
\end{figure*}

Figure~\ref{Fig:Vel} shows the effect of changing the velocity from the model shown in Section~\ref{Sect:BaseModel}: The top row shows an impact at 1~\kms, the middle row shows an impact at 1.5~\kms and the bottom row shows the model shown previously ($v_i$~=~2~\kms). Increasing the velocity increases the peak pressures throughout both the matrix and chondrules. The peak temperatures are significantly higher in the matrix with increasing velocity, while the increase in chondrule temperatures is more modest. As shown in Tables~\ref{Tab:Bulk} and \ref{Tab:Final}, these trends persist for other combinations of initial matrix fraction, porosity and material. The incomplete compaction seen in the reference simulation (top row of Fig.~\ref{Fig:Vel}) is also seen at lower velocities; in these cases, even less pore space is compacted, and the shadow regions are larger. 

Also evident is an increase in chondrule deformation with impact speed (Fig.~\ref{Fig:Vel}). At 1~\kms the chondrules retain their circular shapes whilst at 2~\kms many chondrules are shortened in the direction of shock or inter-chondrule collision.

\begin{figure}[t]
    \centering
        \includegraphics[width=3.4in]{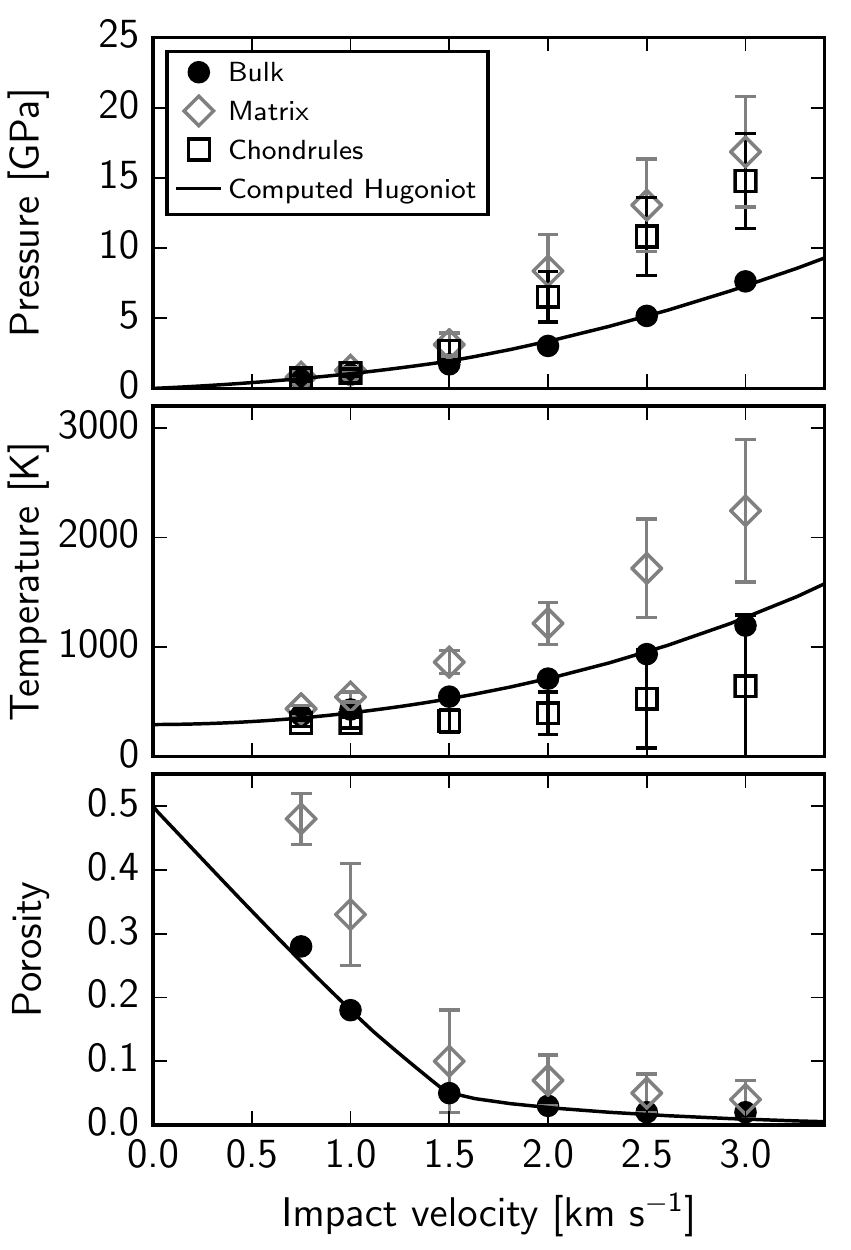}
    \caption{Comparison of the shock pressure (top), peak-shock temperature (middle) and post-shock porosity (bottom) over a range of velocities, for simulations with $A_{mi}$ = 70\% and $\phi_{mi}$ = 0.7. Filled black circles show bulk values of the state quantity, averaged over the entire sample region at a given time (during the shock for temperature and pressure; after release for the porosity); open squares and diamonds show the mean of the state variable in the chondrule and matrix fraction, respectively (error bars show the standard deviation about the mean). The black line shows the computed Hugoniot (see Section~\ref{Sect:eps-alp-params} for details).}
    \label{Fig:BulkResponse}
\end{figure}

Figure~\ref{Fig:BulkResponse} shows how the bulk material and the individual components respond to the shock for a range of velocities from 0.75~\kms to 3~\kms, including the simulations described in this section and Figure~\ref{Fig:Vel} ($A_{mi}$~=~70\%, $\phi_{mi}$~=~0.7). The bulk response of the mixture (solid symbols), calculated as the volume-weighted mean of peak pressure, peak temperature and porosity over all sample tracers, is in good agreement with an estimate of the Hugoniot curve for the bulk material calculated using the equation of state and the $\varepsilon$--$\alpha$ porous compaction model \citep[for details of this calculation, see][]{Davison:10a}. However, the peak pressures and temperatures experienced by the individual components are very different from the bulk response, particularly at high impact speeds. For $v_i \geq$~2~\kms, both the chondrules and matrix see shock pressures 2--4 times higher than those recorded for the bulk material: This is due to a difference in the way the peak and bulk values are calculated. Bulk values are calculated by finding a weighted average across the entire same region \emph{at a given time}. However, peak quantities are recorded as the maximum value that each tracer experienced at \emph{any time throughout the calculation}; thus, resonant oscillations of the shock wave about the steady state may result in peak values that exceed the bulk value. This is in good agreement with behaviour observed in previous mesoscale impact simulations \citep{Gueldemeister:13}.

\subsection{Effect of Initial Matrix Fraction, $A_{mi}$}\label{Sect:Ami}
\begin{figure*}[t]
    \centering
        \includegraphics[width=7in]{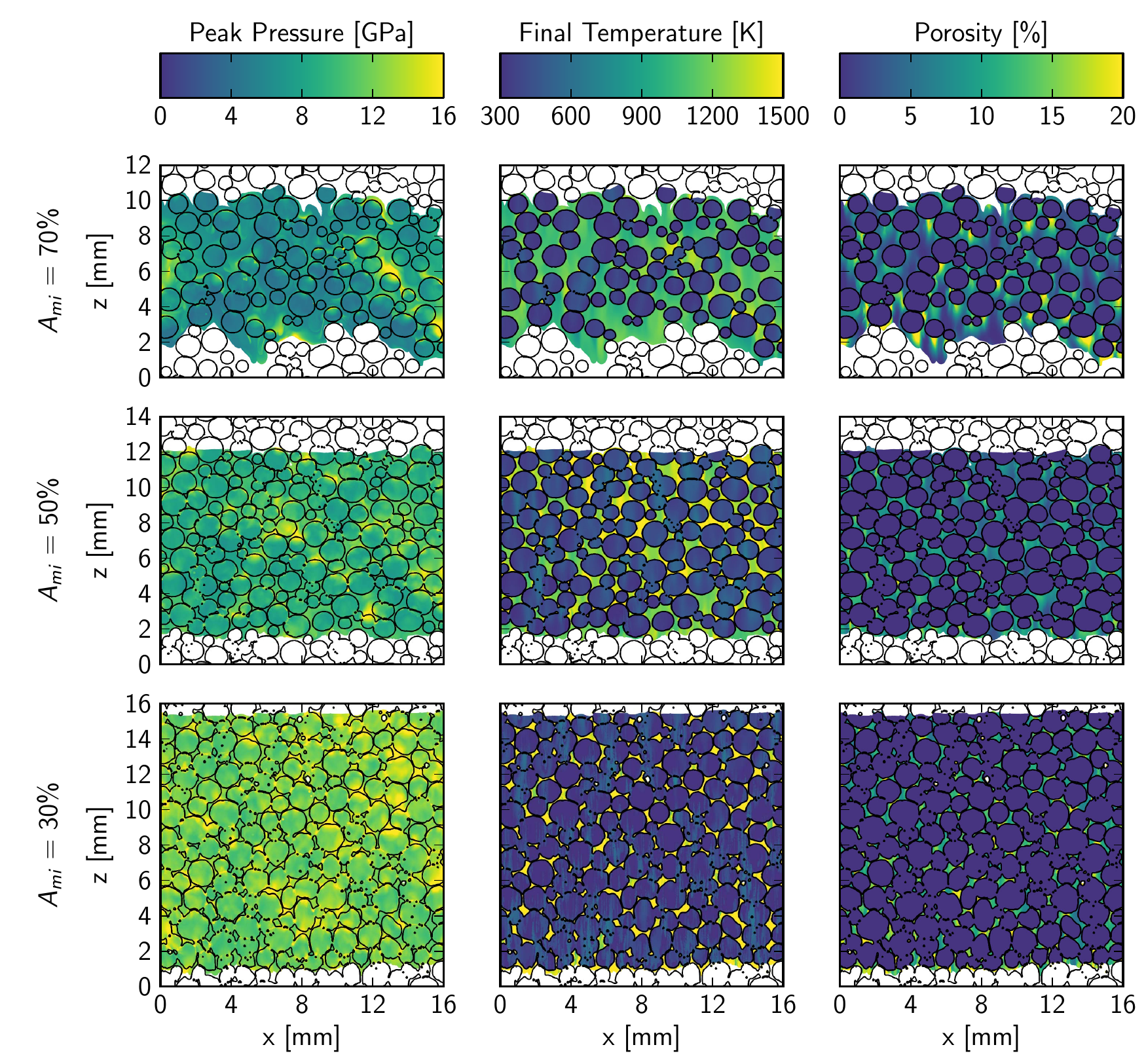}
    \caption{The effect of initial matrix abundance ($A_{mi}$) on the peak pressure, final temperature and porosity.  In these three simulations, the impact velocity ($v_i$ = 2~\kms), matrix porosity ($\phi_{mi}$ = 0.7) and matrix material (dunite) were kept constant.}
    \label{Fig:Abundance}
\end{figure*}

Simulations were run with initial matrix fractions ranging from $A_{mi} = 30\%$ to $A_{mi} = 80\%$. Higher matrix fraction reduces the bulk density and increases the bulk porosity of the impactor and sample material. As shown in Figure~\ref{Fig:Abundance}, this implies  that a decrease in the initial matrix fraction leads to an increase in peak pressures, at fixed impact speed, and consequently higher peak and final temperatures in both the matrix and chondrules. However, the bulk temperature decreases slightly by $\sim$20 K over this range in $A_{mi}$ for $v_i$~=~2~\kms, because the increased volume fraction of cold chondrules reduces the average temperature.

Final matrix porosity appears to be influenced by two competing processes that depend on matrix abundance and impact speed. At low speeds, as matrix abundance decreases, shock pressures are higher, and thus more compaction occurs, leading to lower porosity. However, at velocities high enough for chondrules to come into contact with each other, the porous matrix can be shielded from the shock effects by stress bridges between chondrules, and is compacted less. Chondrule contact becomes more prevalent and occurs at lower impact speeds when the matrix abundance is low. For example, at 1~\kms impact speed, the final matrix porosity decreases with decreasing matrix abundance (compaction dominated), owing to the higher shock pressures that the matrix experiences. At 2~\kms, the final matrix porosity increases as the initial matrix abundance decreases (i.e., there is less compaction when there are more chondrules; stress-bridging dominated), due to the matrix being sheltered in the interstitial spaces between chondrules (bottom left frame of Figure~\ref{Fig:Abundance}). At an intermediate velocity, both of these competing processes (compaction and stress-bridging) can be seen:  At 1.5~\kms, the porosity first decreases with decreasing matrix abundance, and then increases (see Table~\ref{Tab:Bulk}).

\subsection{Effect of Initial Matrix Porosity, $\phi_{mi}$}\label{Sect:Porosity}

\begin{figure*}[t]
    \centering
        \includegraphics[width=7in]{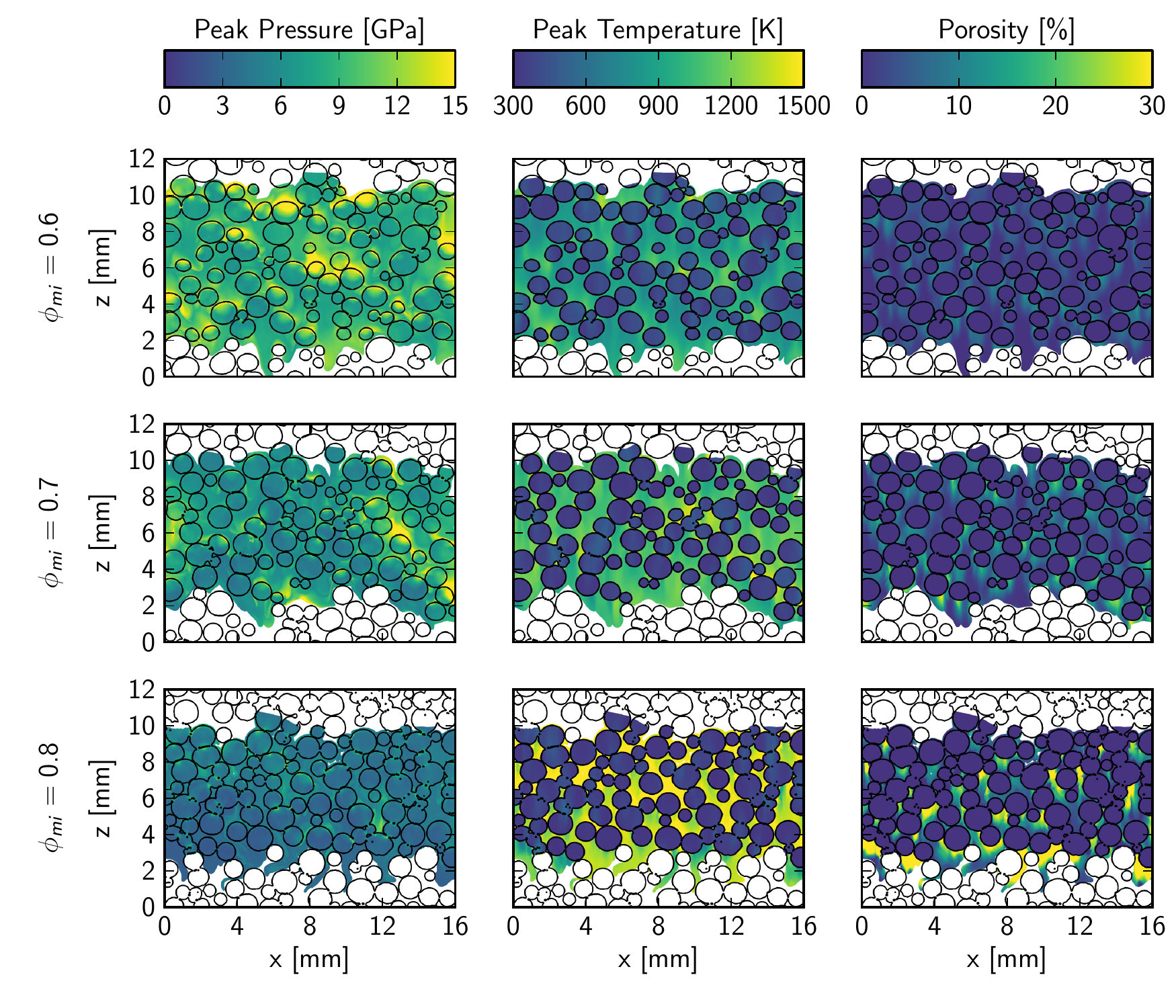}
    \caption{The effect of initial matrix porosity ($\phi_{mi}$) on the peak pressure, peak temperature and porosity. In these three simulations, the impact velocity ($v_i$ = 2 km s$^{-1}$), matrix abundance ($A_{mi}$ = 70\%) and matrix material (dunite) were all kept constant.}
    \label{Fig:Porosity}
\end{figure*}

Figure~\ref{Fig:Porosity} shows the effect of the initial matrix porosity on the peak pressure, peak temperature and final porosity, for impacts at 2~\kms and a matrix abundance of $A_{mi}$ = 70\%. As we would expect, as the initial porosity increases, the peak pressures in both the chondrules and the matrix decrease, due to the reduction in bulk density.
The lower pressures in the non-porous chondrules lead to lower temperatures. However, as the matrix has experienced enhanced compaction (and thus there is additional waste heat in the matrix), the resulting matrix temperatures are higher.
In the cases shown in Figure~\ref{Fig:Porosity}, this leads to an increase in the bulk temperature. With a lower initial matrix abundance, a lower bulk temperature could be the result. The trends in pressure and temperature are the same for the lower velocity simulations at 1~\kms shown in Tables~\ref{Tab:Bulk} and \ref{Tab:Final}. 

\subsection{Effect of Matrix Composition}\label{Sect:Material}

Simulations were also run to test the influence of the choice of material for the matrix. In one set of simulations, the matrix was changed from porous dunite to porous serpentine \citep[using the ANEOS parameters from][]{Brookshaw:98}. As the density of serpentine is lower than dunite (2500~kg~m$^{-3}$ compared to 3314~kg~m$^{-3}$), a second set of serpentine matrix calculations were run, with a matrix porosity of 0.6 (instead of 0.7), to match the bulk density of the matrix in the dunite simulations. Simulations were run at 1, 2 and 3~\kms (Tables~\ref{Tab:Bulk} and \ref{Tab:Final}).

\begin{figure*}[t]
    \centering
        \includegraphics[width=7in]{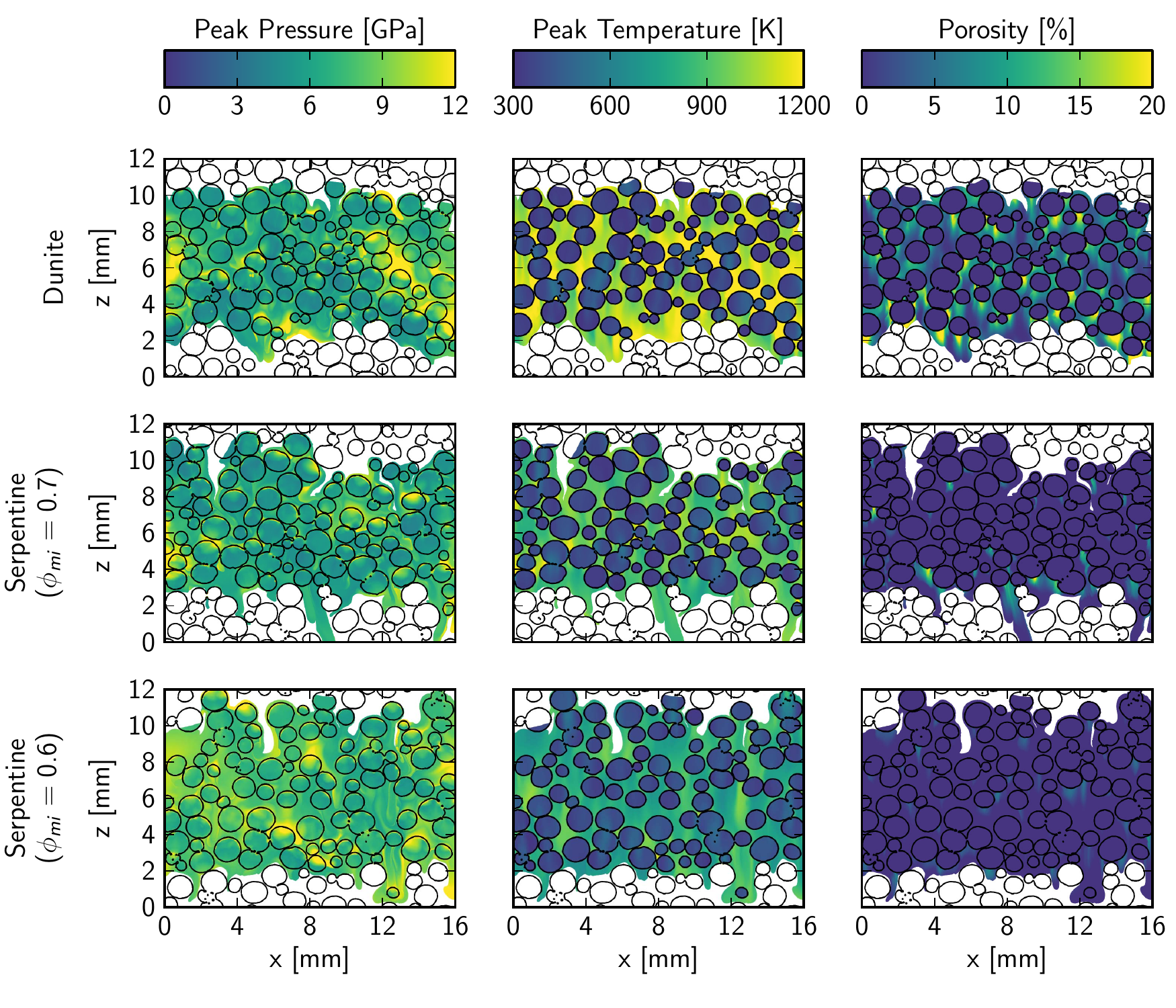}
    \caption{The effect of matrix composition (dunite or serpentine) on the peak pressure, final temperature and porosity. In these three simulations, the impact velocity ($v_i$ = 2~\kms) and matrix abundance ($A_{mi}$ = 70\%) were kept constant.}
    \label{Fig:Composition}
\end{figure*}

At 1~\kms, the pressures and temperatures across all three simulations are similar (although the higher porosity serpentine simulation has slightly higher pressure and temperature than the lower porosity run, as expected from Section~\ref{Sect:Porosity}). The serpentine simulations show more matrix compaction than observed in the dunite simulation. At 2~\kms, the pressure in the matrix is around 0.5~GPa higher in the $\phi_{mi} = 0.6$ serpentine model than in the dunite model, and around 0.7~GPa lower in the  $\phi_{mi} = 0.7$ porosity serpentine model than in the dunite model. However, in both cases, the temperatures in the matrix are significantly lower than in the dunite simulation (814~K and 941~K in the 60\% and 70\% serpentine, compared to 1110~K in the dunite; Figure~\ref{Fig:Composition}). The same observation can be made for the 3~\kms simulations.  The low serpentine matrix temperature is a consequence of the phase change of the water content buffering the temperature increase. 

\section{DISCUSSION}\label{Sect:Discussion}

\subsection{Determination of $\varepsilon-\alpha$ Model Parameters For Use in Macroscale Simulations}
\label{Sect:eps-alp-params}
The $\varepsilon-\alpha$ porous compaction model \citep{Wuennemann:06,Collins:11a} is used to parameterise the bulk response of porous materials to shock compaction in impact simulations. The input parameters for this model are typically constrained by experimental data; however, they can also be determined using the results of mesoscale compaction simulations. The $\varepsilon-\alpha$ model was developed as an alternative to the conventional $P-\alpha$ model \citep{Hermann:69}, because it has the efficiency advantage of computing the distension ($\alpha$) directly from the volume strain ($\varepsilon$). As volume strain is usually computed before the pressure in shock physics models such as iSALE, this avoids the need for an iteration to find pressure and distension simultaneously. There are four regimes in the $\varepsilon-\alpha$ model to describe the compaction and compression of a porous material: elastic compaction, exponential compaction, power-law compaction and compression. Permanent compaction of pore space occurs in the exponential and power-law compaction regimes, which are defined, respectively, as: 

\begin{eqnarray}
    \alpha & = & \alpha_0e^{\kappa(\varepsilon-\varepsilon_e)} \label{Eq:ExpCompaction}\\
    \alpha & = & 1 + (\alpha_x-1)\left(\frac{\varepsilon_c-\varepsilon}{\varepsilon_c-\varepsilon_x}\right)^2 \label{Eq:PowerCompaction}
\end{eqnarray}

where $\alpha_0$ is the initial distension, $\alpha_x$ is the distension at the transition from the exponential regime to the power-law regime, and $\kappa$ is the compaction rate parameter. There are three volume strains defined in Equations~(\ref{Eq:ExpCompaction})~\&~(\ref{Eq:PowerCompaction}) which correspond to the transitions between the four regimes listed above: $\varepsilon_e$ is the volume strain at the transition from the elastic (reversible) to exponential (irreversible) compaction regimes. For high porosity materials, $\varepsilon_e$ is small and has a very minor effect on the compaction curve; hence, in this work $\varepsilon_e$ was set to a constant $-10^{-5}$ for all porous mixtures. $\varepsilon_x$ corresponds to the transition from the exponential to the power-law compaction regime, and is derived by setting $\alpha=\alpha_x$ in Equation~(\ref{Eq:ExpCompaction}). $\varepsilon_c$ is the volume strain at which all pore space is crushed out (at the transition from power-law compaction to pure compression). It is derived by differentiating Equation~(\ref{Eq:PowerCompaction}), and setting $d\alpha / d\varepsilon = 0$ to ensure a smooth termination of compaction. Thus, the bulk permanent compaction of our mesoscale mixtures can be characterised by two material specific parameters ($\alpha_x$ and $\kappa$). Here we derive estimates for the values of those two parameters by fitting Hugoniot curves calculated using the $\varepsilon$-$\alpha$ model to the mesoscale simulation results for a given initial matrix fraction and matrix porosity over a range of impact velocities.

\begin{figure*}
    \includegraphics[width = 7in]{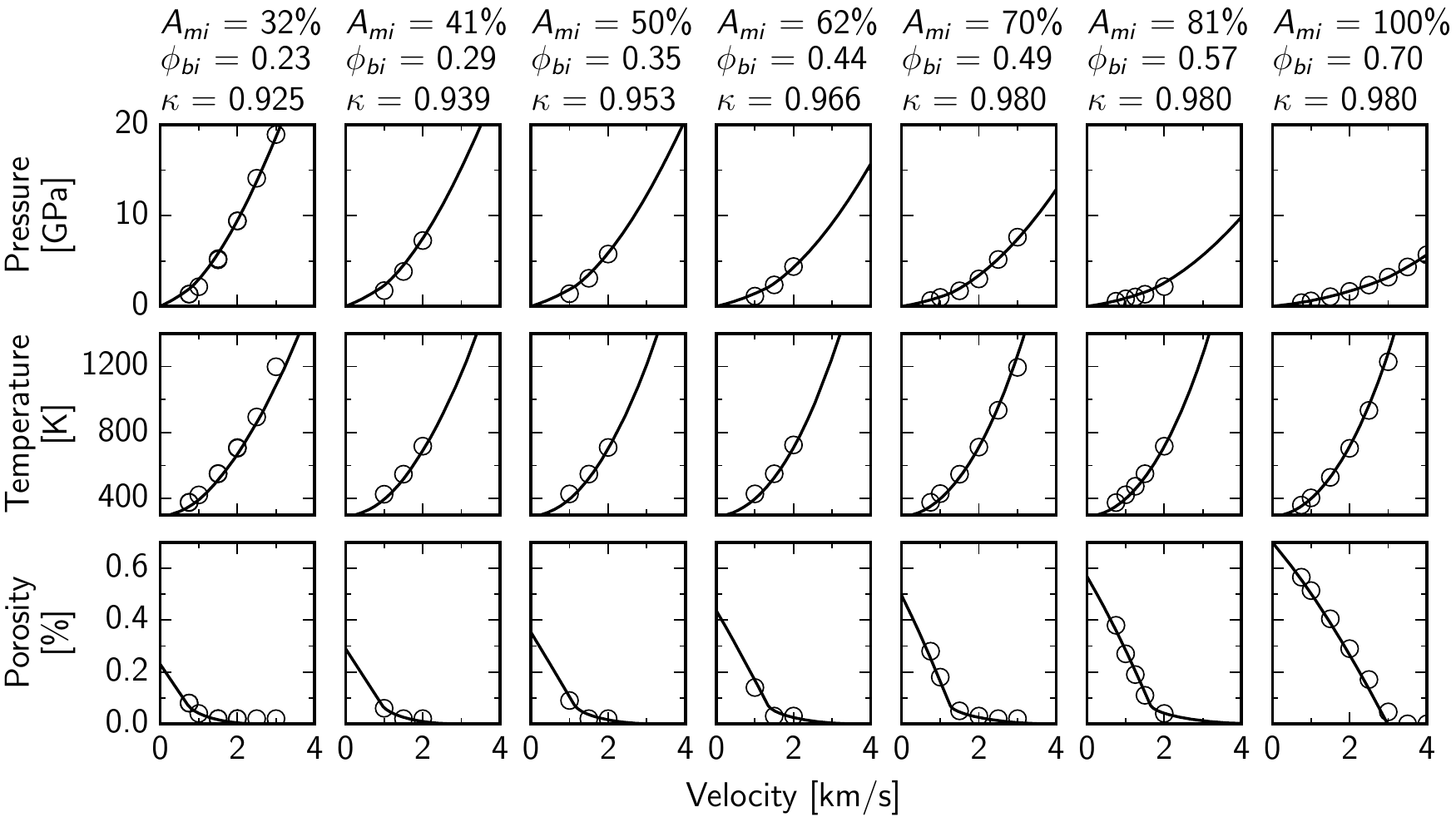}
\caption{Hugoniots for a range of bimodal mixtures in the pressure-impact velocity (top row), temperature-impact velocity (middle row) and porosity-impact velocity (bottom row) planes. Open circles show the values output from iSALE simulations (see Table~\ref{Tab:Bulk}), and solid lines show the hugoniot computed from the equation of state and $\varepsilon-\alpha$ porous compaction model \citep{Davison:10a}.}
\label{Fig:epsalp}
\end{figure*}

For the three mixtures which have simulations with five or more impact velocities ($A_{mi}$ = 32\%, 70\% and 81\%), we find that $\alpha_x$ = 1.06 fits all three well; however the value of $\kappa$ is dependent on the bulk porosity. For $A_{mi}$ = 70\% and 80\% (i.e. a bulk porosity of 0.49 and 0.57, respectively), $\kappa$ = 0.98 fits the simulated data well \citep[this agrees well with previous simulations of $\sim$ 50\% porous material; for example][]{Wuennemann:08,Davison:10a}. This is shown in Figure~\ref{Fig:epsalp}. For $A_{mi}$ = 32\% ($\phi_{bi}$ = 0.23), $\kappa$ = 0.925, a value somewhat lower than used before. 

Using $A_{mi}$ = 32\% and $A_{mi}$ = 70\% as endpoints, the value of $\kappa$ for $A_{mi}$ = 41\%, 50\% and 62\% was estimated using a linear interpolation ($\kappa$ = 0.939, 0.953 and 0.966, respectively), and was found to fit the iSALE data well (Figure~\ref{Fig:epsalp}). Thus, for future macroscale simulations of planetesimal collisions, the $\varepsilon-\alpha$ model parameter $\kappa$ can be determined using the following relationship:

\begin{equation}
\kappa =   0.0014 A_{mi} +   0.88
\end{equation}

For the given matrix porosity used here ($\phi_{mi}$ = 0.7), that corresponds to the following relationship:

\begin{equation}
\kappa =  0.21 \phi_{bi} + 0.88
\end{equation}

\subsection{Equilibration Timescale}
\label{Sect:EquilibrationTime}

The temperature dichotomy between the matrix and chondrules observed in all mesoscale simulations will be short lived, as the cold chondrules act as heat sinks inside the heated matrix. To determine the approximate timescale of this equilibration, a 1-D finite difference calculation was performed to solve the heat conduction equation:
\begin{equation}
C_p\left(x\right) \rho\left(x\right) \frac{\partial T}{\partial t} = \frac{\partial}{\partial x} \left( K\left(x\right) \frac{\partial T}{\partial x}\right)
\end{equation}
where $C_p$ is specific heat capacity, $\rho$ is density, $K$ is conductivity, $T$ is temperature, $t$ is time and $x$ is distance. %
\subsubsection{Simulation Design and Initial Conditions}
$C_p$ was set to a constant value of 800~J~kg$^{-1}$~K$^{-1}$ \citep[e.g.][who found that to be a good match to solar system materials]{Davison:12}, $\rho$ was the density of the two components ($\rho_0$~=~3314~kg~m$^{-3}$ for the non-porous chondrules), and modified for the matrix using $\rho=\rho_0(1-\phi_{mf}$). $K$ was also assigned with a dependence on porosity; values of diffusivity ($\kappa$) have been reported for meteoritic materials previously, where diffusivity $\kappa = {K}/{C_p \rho}$, ranging from $1 \times 10^{-7}$~m$^2$~s$^{-1}$ \citep{Ghosh:99,Ghosh:03} to $\sim 7 \times 10^{-7}$~m$^2$~s$^{-1}$ \citep{Opeil:10}. This gives two end-member conductivity values  for the non-porous material of $K_0 =$ 0.27  and 1.86~W~m$^{-1}$~K$^{-1}$, respectively. The conductivity is then modified for the porous matrix using the scaling relationship from \citet{Warren:11}:
\begin{equation}
K = K_0 e^{-12.46\phi}
\end{equation}
In order to turn the information from the 2D iSALE simulations into a 1D problem here, the calculations were initialised as follows: The mesh was divided into two parts --- chondrule and matrix. The location of the boundary was determined by the final matrix-to-chondrule volume ratio ($A_{mf}$); for example, if $A_{mf} = 40\%$, the mesh would be composed of 60\% chondrule and 40\% matrix. The temperature of each component was set to the mean post-shock temperature for that component from the iSALE simulation ($T_{cf}$ and $T_{mf}$; {Table~\ref{Tab:Final}), and the porosity of the matrix was assigned the mean post-shock matrix porosity ($\phi_{mf}$; Table~\ref{Tab:Bulk}). The calculation was allowed to continue until the standard deviation of the temperature field had decreased below a threshold value, here taken to be 1\% of the bulk temperature.%

\subsubsection{Results}
This calculation was performed for the six simulations listed in Tables~\ref{Tab:Bulk} and \ref{Tab:Final} with $A_{mi} = 70\%$ and $\phi_{mi} = 0.7$ (with velocities ranging from 0.75~km~s$^{-1}$ to 3~km~s$^{-1}$). The time taken for these simulations to equilibrate to the bulk temperature are presented in Table~\ref{Tab:Equilibrate}. As shown in Figure~\ref{Fig:Equilibrate}, the equilibration time is strongly dependent on the choice of $\kappa_0$. As we would expect, the time taken when $\kappa_0 = 7 \times 10^{-7}$~m$^2$~s$^{-1}$ is a factor of 7 shorter than the time taken when $\kappa_0 = 1 \times 10^{-7}$~m$^2$~s$^{-1}$. As the diffusivity in the matrix is also controlled by the porosity, there is a clear dependence of equilibration time on matrix porosity. For cases where the matrix porosity is 0.1 or less, the equilibration time is on the order of a less than a second to a few seconds. For higher porosity ($\phi_{mf}$~=~0.3--0.5), the equilibration time takes 10's to 100's of seconds.

\begin{table}
    \begin{center}
    \caption{Timescale for equilibration of temperature dichotomy after the shock event.} 
    \label{Tab:Equilibrate}
    \begin{tabular}{cccc}
        \hline\hline
        & & \multicolumn{2}{l}{Equilibration timescale (s)} \\[4pt]
        \cline{3-4}\\[-8pt]
        Impact velocity  & Matrix  & \multicolumn{2}{l}{$\kappa_0$ (m$^2$~s$^{-1}$)}\\
        (km~s$^{-1}$) & porosity & $1 \times 10^{-7}$ & $7 \times 10^{-7}$\\[4pt]
        \hline
        0.75 & 0.48 & 690 & 98  \\
        1.00 & 0.33 & 110 & 16  \\
        1.50 & 0.10 & 6.3 & 0.89 \\
        2.00 & 0.07 & 5.0 & 0.71 \\
        2.50 & 0.05 & 4.0 & 0.57 \\
        3.00 & 0.04 & 4.0 & 0.58 \\
        \hline
    \end{tabular}
    \end{center}
\end{table}

\begin{figure}
    \centering
    \includegraphics[width=3.4in]{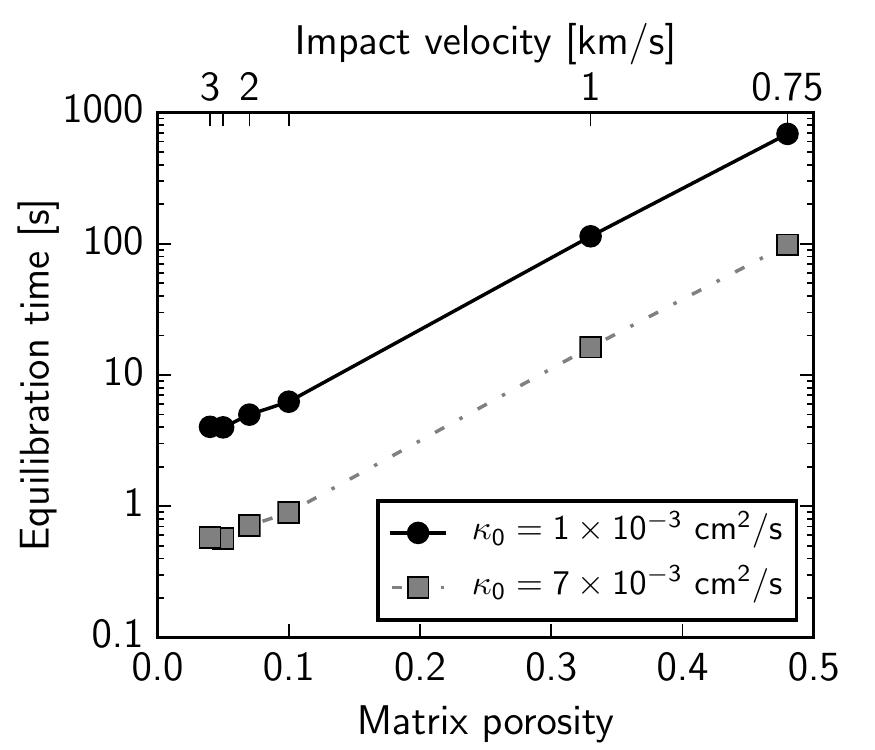}
    \caption{Timescale for chondrule and matrix temperature to equilibrate in the suite of simulations with $A_{mi} = 70\%$ and $\phi_{mi} = 0.7$.}
    \label{Fig:Equilibrate}
\end{figure}

An estimate for the length of time a material will remain under high pressure for during a shock event is given by $L / U_p = 2L / v_i$ \citep[where $L$ is ths impactor diameter, $U_p$ is the particle velocity and $v_i$ is the impact velocity;][]{Melosh:89}. For the timescales presented in Table~\ref{Tab:Equilibrate} and Figure~\ref{Fig:Equilibrate}, this would mean that for impactor diameters of $>$~10~km, it is possible that the equilibration timescale would be shorter than the shock duration (and thus suggests equilibration may be complete before the arrival of the release wave). Since most impacts on meteorite parent bodies come from impactors smaller than 10~km \citep{Davison:13}, this should not be an important effect for most impacts. Further modeling to understand the effects of equilibration before release is required, but is beyond the scope of this article.
\subsection{Meteorite Groups} 
\label{Sect:MeteoriteGroups}
Different meteorite groups have different ranges of bulk porosities and matrix fractions. For example, carbonaceous chondrites have a range of porosities of $\sim$~0.04--0.28 \citep{Macke:11} and matrix abundances of 30--70\% \citep{Scott:03}, while ordinary chondrites have porosities in the range $\sim$~0.06--0.16 \citep{Consolmagno:98} and matrix abundances in the range 10--15\% \citep{Scott:03}. By choosing simulations with different starting conditions and impact velocities, final compositions can be produced which are similar to those observed in the different meteorite groups. Figure~\ref{Fig:MeteoriteGroups} shows the final bulk porosities and matrix fractions for a range of simulations with different velocities, for two different starting conditions. The similarity between simulations that started with $A_{mi} = 70\%$ and carbonaceous chondrites, and between simulations with $A_{mi} = 30\%$ and unequilibrated ordinary chondrites shows that the compaction processes presented in this work are able to reproduce final properties consistent with real meteorite observations (Figure~\ref{Fig:MeteoriteGroups}(b)).

\begin{figure}[t]
    \centering
    \includegraphics[width=3.4in]{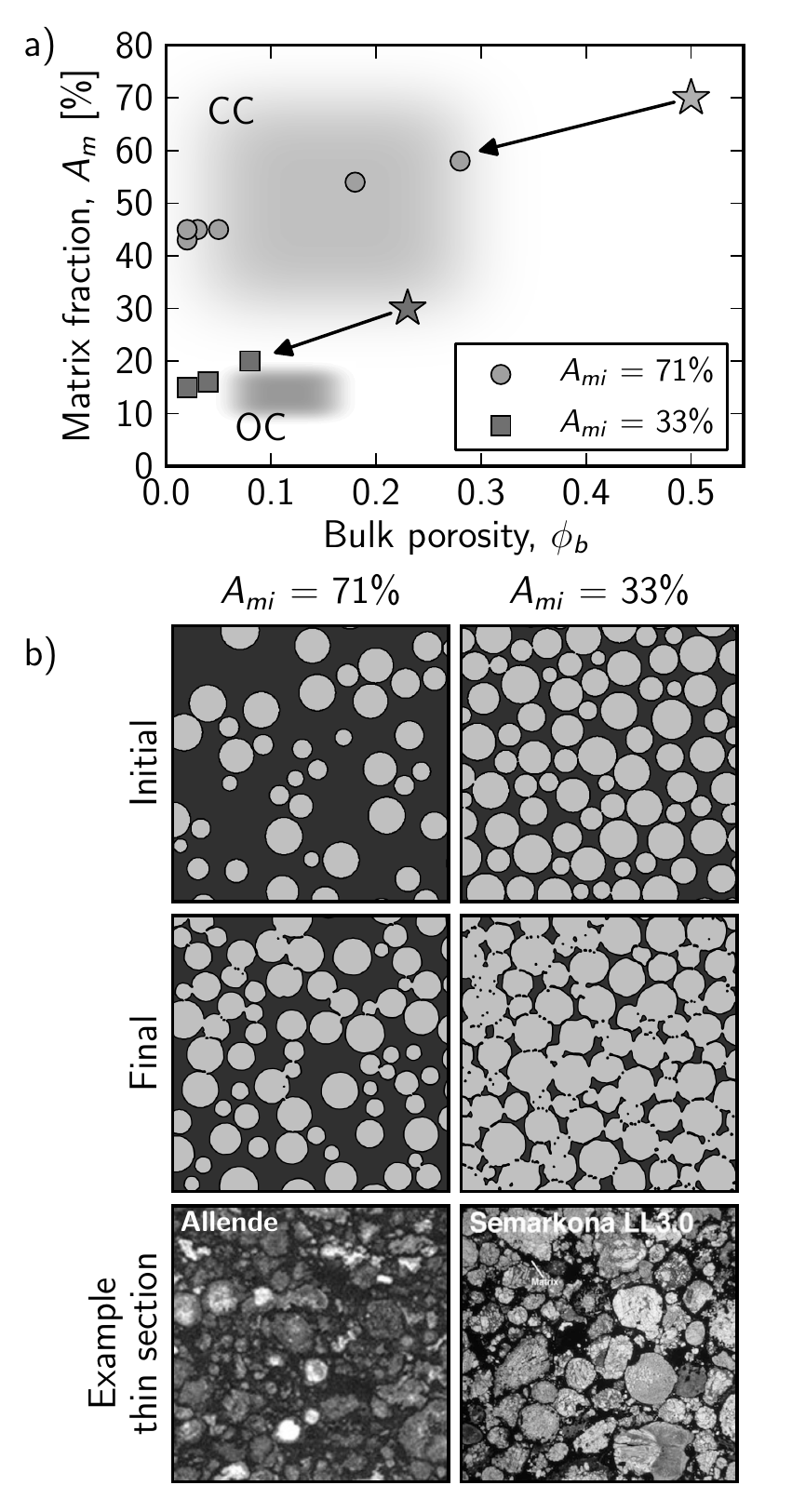}
    \caption{(a) Final matrix fraction and final bulk porosity from a range of impact simulations with two different starting matrix fractions, denoted by stars. Final condition is dependent of velocity, with higher velocities leading to more compaction, and thus lower matrix fractions (and, by extension, lower bulk porosity). Shaded rectangles show the range of typical values for carbonaceous chondrites (CC) and ordinary chondrites (OC) \citep{Consolmagno:98,Scott:03,Macke:11}. (b) Initial and final distribution of chondrules (light grey) and matrix (dark grey) for two simulations at 1.5~\kms with initial matrix fractions of 70\% and 30\%. The final states are comparable to thin sections of Allende (a carbonaceous chondrite) and Semarkoma (an ordinary chondrite). Allende image reprinted from \citet{MacPherson:11}, with permission. Semarkona image reprinted from \citet{Weisberg:06}, from Meteorites and the Early Solar System by D. S Lauretta and H. Y. J. McSween. The Arizona Board of Regents \copyright 2006. Reprinted by permission of the University of Arizona Press.}
    \label{Fig:MeteoriteGroups}
\end{figure}

\subsection{Model Limitations} 
The simulations described above are a significant first step in quantifying the heterogeneous response of primordial solar system solids to shock compaction. However, there are several limitations of this approach which should be addressed in future. 

One key limitation is the use of 2D plane-strain geometry, rather than a more realistic 3D geometry. This assumption was necessary for this large-parameter-space study to limit computational cost (both in terms of time and memory).  In a simulation with 3D geometry, out of plane contacts between chondrules would likely stiffen the bulk response of the mixture (compared to the plane-strain geometry), particularly in scenarios where the initial chondrule volume fraction is high. However, based on similar numerical mesoscale studies of pore-space compaction \citep{Gueldemeister:13} in which in 2D and 3D geometries were simulated, we expect qualitatively similar behaviour in both geometries: The magnitude and length-scale of the pressure-temperature heterogeneity and the trends in the heterogeneity with impact velocity, initial chondrule volume fraction, matrix porosity and matrix material should all be qualitatively similar.

The $\varepsilon-\alpha$ porous compaction model and the equation of state tables used to describe both the chondrules and solid-component of the porous matrix (dunite/forsterite and serpentine) are also over-simplified. First, the compaction model assumes that all of the $PdV$ work deposited by the shock in the porous matrix will lead to a temperature increase. In reality, dissipative processes during compaction, such as grain deformation and fracturing will lead to an increase in entropy as well as temperature. Neglecting the entropy increase during crushing will result in an overestimate of shock heating, but this is difficult to quantify without experimental measurements of shock heating. Second, the version of ANEOS that we used to derive the forsterite and serpentine tables does not permit both solid-solid and solid-liquid phase transitions to be included at the same time \citep[e.g.][]{Melosh:07}. As in previous work \citep[e.g.][]{Davison:10a} we regarded the effect of the solid-liquid phase transition as less important than that of the solid-solid phase transition. Neglecting latent heat of melting implies that temperatures in the table that exceed the solidus are over-estimated. 

At higher shock pressures, recent shock compression experiments of quartz \citep{Hicks:06} suggest that ANEOS over estimates the temperature increase and under estimates the entropy increase during shock compression, because it assumes a heat capacity in the fluid region that is too low \citep{Kraus:12}. If this limitation of ANEOS is also important for other silicate rocks, it implies that the shock pressure required to vaporise the matrix is over-estimated by ANEOS and that peak and post-shock temperatures above the liquidus are also over-estimated. The primary focus of the simulations in this work was relatively low velocity collisions, which in general produced matrix heating below and up to the solidus; thus, this limitation of ANEOS is of minor significance to our conclusions. Future simulations in which higher-velocity collisions are considered will need to address this limitation. Work is underway to produce ANEOS tables that include two phase transitions \citep[e.g. solid-solid and solid-liquid,][]{Collins:14}, which will be important in those cases.

For the reasons described above, the temperatures (both peak and post-shock) quoted in Tables~\ref{Tab:Bulk} and \ref{Tab:Final}, and particularly those above the solidus, should be considered as upper limit for each given impact scenario. However, since the peak and post-shock temperatures are strongly dependant on the initial matrix porosity and initial temperature, using less conservative initial conditions (e.g., higher initial porosity or starting temperature) in our models could easily compensate for any overestimate in temperature due to inadequacies of the material model. Thus, while the exact temperatures in a given simulation may change with initial conditions or model assumptions, the relative trends of increasing temperature with impact velocity, matrix porosity and chondrule volume fraction are robust.

\section{CONCLUSIONS}
In this work, we adapted a mesoscale modeling technique and applied it to meteoritic material. We investigated the effects of impact velocity, composition, matrix porosity and abundance on the shock processing of meteoritic material during an impact. Our results show that low-speed impact compaction of chondrule-matrix mixtures can reproduce observed properties of meteoritic samples. 
One key observation from our simulations is a strong temperature dichotomy between the chondrules (cold) and the matrix (hot) after an impact, even at relatively low speeds. This temperature dichotomy will be short lived (seconds to minutes), and the timescale of equilibration is dependant on the post-shock porosity of the matrix --- since the thermal diffusivity decreases with increasing porosity, more porous materials will stay hot for longer. 
In addition to the strong dichotomy in the temperature between the matrix and the chondrules, we have observed heterogeneous heating and compaction within the matrix itself; for example, on the lee side of chondrules the matrix may be protected from the shock and thus experience less compaction than the surrounding matrix.

In materials with a low matrix abundance (i.e. more chondrules), as more chondrules come into contact with each other (at velocities of $>$~2~\kms), the matrix can be protected by the chondrules forming stress bridges, resulting in less compaction than that observed in higher matrix abundance materials.
At low impact velocities ($<$~2~\kms), chondrules typically remain circular. However, at velocities of 2~\kms and above, many chondrules are deformed; they become shortened in either the direction of the shock or due to inter-chondrule collisions.

Using the mesoscale simulation results, we have constrained $\varepsilon-\alpha$ porous compaction model parameters as a function of matrix abundance or bulk porosity, which are appropriate for describing the bulk material response in macroscale simulations of planetesimal collisions.

\acknowledgments
TMD and GSC were funded by STFC grant ST/J001260/1. PAB acknowledges the support of the Australian Research Council via the Australian Laureate Fellowship scheme. We thank the developers of iSALE (www.isale-code.de), and are grateful to Fred Ciesla for discussions about the equilibration timescale calculations. We thank the anonymous reviewer for their valuable comments.\\

\appendix

\section{A NEW METHOD FOR MOVING LAGRANGIAN TRACER PARTICLES}
\label{Apdx:Tracers}
The techniques presented in this work for simulating the mesoscale effects of shockwaves on meteoritic material require the use of Lagrangian tracer particles, which allow the history of a particular parcel of material to be tracked as it moves through the fixed (Eulerian) mesh in iSALE. Here we describe a new method implemented in iSALE for moving tracer particles. 

The original method for displacing a tracer in iSALE is a simple forward Euler projection of the tracer position using the interpolated velocity vector at the tracer location multiplied by the current timestep duration. Velocity interpolation is bilinear in the $x$ and $y$ directions, using the velocity stored at the four nodes of the cell within which the tracer is located at the start of the timestep.

If $\xi^n_{x}$ and $\xi^n_{y}$ are the fractional distances across the cell of the tracer from the four cell nodes ($n$ = 1--4, counter-clockwise from the bottom left), and $v^n_{x}$ and $v^n_{y}$ are the horizontal and vertical components of the velocity at the node of the cell, then the horizontal and vertical components of the tracer's velocity are calculated as such:

\begin{equation}
    (v_x,v_y)  = \sum_{n=1}^{4} (1-\xi^n_{x})(1-\xi^n_{y}) (v^n_x,v^n_y) \label{Eq:vxy}
\end{equation}

This method works well in simulations with few material boundaries, and thus few cells containing multiple materials (typical iSALE simulations involve an impactor into a target with up to three layers, and thus fewer mixed cells than the simulations presented in this work). However, as this approach does not respect material boundaries it often results in tracers drifting away from the material they were intended to track through the simulation (see the top panel of Figure~\ref{Fig:Tracers}). This problem is more severe in simulations with many material interfaces and a high proportion of mixed cells. To overcome this issue, a new method was developed for this work, in which tracers are moved according to material fluxes into and out of each face of the cell in which they are positioned. This information is already calculated during the advection step of the iSALE cycle and so does not entail a significant computational overhead. 

Using the fractional distances from the bottom left of the cell (i.e. $\xi^1_{x}$ and $\xi^1_{y}$), the horizontal and vertical velocity components, $v_x$ and $v_y$, can be calculated as:\\

\begin{equation}
    v_x = \left(\frac{(1 - \xi^1_{x})F_L}{\alpha_L} - \frac{\xi^1_{x} F_R}{\alpha_R}\right)\frac{x_{i+1} - x_i}{V_{i,j}\ dt}\label{Eq:vx2} 
\end{equation}
and

\begin{equation}
    v_y = \left(\frac{(1 - \xi^1_{y})F_B}{\alpha_B} - \frac{\xi^1_{y} F_T}{\alpha_T}\right)\frac{y_{j+1} - y_j}{V_{i,j}\ dt}\label{Eq:vy2} 
\end{equation}
where $F$ is the net volume of the tracer's material fluxing into the cell through each face (subscripts $L, R, B$ and $T$ denote the left, right, bottom and top faces, respectively), $\alpha$ is the volume fraction of the tracer's material fluxing through the face compared to the total volume of material (i.e. $F_m/F_{tot}$), $V_{i,j}$ is the volume of the cell $i, j$, and $dt$ is the length of the current time step.

To ensure numerical stability, in cases where the volume fraction of the tracer's material is 0 on one face of the cell, the tracer is assumed to be on the opposite face and the term for that face in Equations~\ref{Eq:vx2} or \ref{Eq:vy2} is ignored; this avoids a divide-by-zero error. For example, if $\alpha_L = 0$, then $s_x$ is assumed to be 1, and the term ${(1 - s_x)F_L}/{\alpha_L}$ is set to 0.

Finally, some checks must be made to ensure that the velocity calculated here is not going to (a) move the tracer into a cell that does not contain any of the tracer's material, or (b) leave the tracer in a cell that will become empty of the tracer's material during the timestep. These two cases are countered by finding the tracer's distance from the cell face, and then moving the tracer along its current trajectory (determined by the velocity components in Equations~\ref{Eq:vx2} and \ref{Eq:vy2}) far enough to just cross the cell face. For case (a) above, this will return the tracer to its original cell, and for case (b) it will move the tracer into the cell where the material is also moving.

As shown in Figure~\ref{Fig:Tracers}, this means that tracers remain with their material, and do not get isolated from their material as was possible in the prior cell-node velocity method. Note that in Figure~\ref{Fig:Tracers} where some tracers appear to have crossed the material boundary in the material method, they are in mixed-material cells (i.e. a cell with a material boundary in it) and thus are still attached to their respective material --- this is just a result of how the material boundary contours were drawn while constructing the figure.

We note that the problem of tracer drift that we ameliorate with the above algorithm is a consequence of material being advected through the mesh at a different speed to the bulk flow. This, in turn, is a consequence of the multi-material advection and interface construction algorithms in iSALE, which modify material fluxes between cells to preserve sharp interfaces between materials. In other words, tracer drift is not necessarily a limitation of iSALE's original method for moving tracers and the optimum tracer movement algorithm for a given problem will depend on whether it is more important to track material history or kinematics. The new method presented here is preferable when a faithful record of material state through the simulation is a priority.

\end{document}